\documentclass[11pt, oneside]{article}   	
\usepackage[margin=1in]{geometry}               
\usepackage{graphicx}				
										
\usepackage{amssymb}
\usepackage{physics}
\usepackage{amsmath}
\usepackage[dvips]{epsfig}

\usepackage{amsfonts}
\usepackage{subfig}

\usepackage{mathrsfs}

\usepackage{xcolor}

\usepackage{bm}

\usepackage{blindtext}

\usepackage[title]{appendix}

\usepackage{cite}

\usepackage{authblk}

\def\cchi{\raise2pt\hbox{$\chi$}} 

\newcommand\bsdot{\ensuremath{\boldsymbol{.}}}

\title{\bf{Two Dimensional Electromagnetic Scattering from Dielectric Objects using Qubit Lattice Algorithm}}

\author{George Vahala}
\affil{Department of Physics, William \& Mary, Williamsburg, VA23185}
\author{Min Soe}
\affil{Department of Mathematics  and Physical Sciences, Rogers State University, Claremore, OK 74017}
\author{Linda Vahala}
\affil{Department of Electrical \& Computer Engineering, Old Dominion University, Norfolk, VA 23529}
\author{Abhay K. Ram}
\affil{Plasma Science and Fusion Center, MIT, Cambridge, MA 02139} 

\begin{document}
\maketitle
$\bf{Abstract}$: A qubit lattice algorithm (QLA) is developed for Maxwell equations in a two-dimensional Cartesian geometry.
In particular, the initial value problem of electromagnetic pulse scattering off a localized 2D dielectric object is considered.
A matrix
formulation of the Maxwell equations, using the Riemann-Silberstein-Weber vectors, forms a basis for the QLA and a possible unitary representation. The electromagnetic
fields are discretized using a 16-qubit representation at each grid point. The discretized QLA equations reproduce Maxwell equations to
second order in an appropriate expansion parameter $\epsilon$.  The properties of scattered waves depend on the scale length
of the transition layer separating the vacuum from the core of the dielectric. The time evolution of the fields gives interesting
physical insight into scattering when propagating fields are excited within the dielectric medium. Furthermore, simulations
show that the QLA recovers Maxwell equations even when $\epsilon \sim 1$.



\section{Introduction}

\qquad It is important to develop algorithms that can be ideally parallelized on classical supercomputers and,
with a little tweaking, can also be run on error-correcting quantum computers.  One class of such mesoscopic
algorithms is the qubit lattice algorithm (QLA) [1-20].  QLA typically consists of an interleaved sequence of unitary collision
and streaming operators acting on an appropriate basis set, e.g., a basis set of qubits.
Under appropriate moments and in the continuum limit the QLA operators can be so chosen that the desired continuum
  equations will be recovered.
For example, QLAs have been developed for the one-dimensional (1D) nonlinear Schrodinger equation (NLS) and
the Korteweg-De Vries (KdV) equation [6, 9-11] to study soliton dynamics and to benchmark the QLA.
 Different choice of unitary collision operators are needed to, for example, recover the $\partial^2/\partial x^2$ for NLS or remove this second derivative but have instead the $\partial^3/\partial x^3$ for KdV.

The application of quantum information science to plasma physics has drawn interest in recent years [21-24].
Electromagnetic waves are ubiquitous in laboratory, space, and astrophysical plasmas. The waves can be externally generated by an
antenna and coupled to a plasma, or internally generated by a plasma instability. In either case, an analysis of the propagation and damping of
waves is essential -- whether for determining the spatial profile of power deposition or for diagnosing the interior of a hot plasma.
The allure for developing a QLA for electromagnetic wave propagation is the connection between the relativistically invariant Maxwell equations
and the Dirac equation for a relativistic particle [25-27]. Following this connection, Khan [28] has used the Riemann-Silberstein-Weber vectors
to express Maxwell equations, for an inhomogeneous scalar dielectric medium, as a matrix evolution equation of an 8-spinor. 
However, in this representation,  the 8-spinor evolution equation is not fully unitary; the operators which include the derivative of the refractive index are Hermitian.
For a homogeneous dielectric medium, the 8-spinor representation decouples into two independent 4-spinor equations which are immediately unitary.  (Physically this implies that in a homogeneous medium the two electromagnetic field polarizations decouple.)

The backbone of the QLA for Maxwell equations is the realization that, for a homogeneous medium, the 4-spinor representation has
the same  structure as the Dirac equation for a massless free particle[1, 17].  The only difference being in the \textcolor{black}{detailed} linear couplings of the spinor
components. Thus, we are able to generalize Yepez's [1] QLA construction for the Dirac equation to construct the required unitary collision
operator which recovers Maxwell equations to second order accuracy in terms of a small parameter introduced in the collision angle.   

Recently [17, 29] we studied the 1D QLA for Maxwell equations by considering the initial value problem of a \emph{pulse} propagating normal
to a dielectric slab.  This is a generalization of the classic boundary value problem of an incident \emph{plane wave} onto a discontinuous dielectric medium.  From the boundary conditions at the interface, assuming reflected and transmitted plane waves, one obtains the classic Fresnel equations for
the ratios of the electric field amplitudes,
\begin{gather*}
\frac{E_{refl}}{E_{inc}	} = \frac{n_1 -n_2}{n_1 + n_2} \qquad , \qquad  \frac{E_{trans}}{E_{inc}	} = \frac{2 n_1}{n_1 + n_2}
\end{gather*}
where the plane wave propagates from medium with refractive index $n_1 \rightarrow n_2$.  In our 1D QLA initial value problem of an incident pulse onto a thin dielectric boundary layer we immediately realize that there are significant differences from the boundary value plane wave problem:  our pulse has finite width and at particular time intervals  it will transiently \textcolor{black}{straddle} the boundary layer region with parts of the pulse in medium $n_1$ and in medium $n_2$.   \textcolor{black}{During these time intervals the pulse shape is very different from the initial pulse geometry.}
In the QLA simulations no internal boundary conditions are applied, 
\textcolor{black}{and there are no assumptions on the existence or geometric form of the scattered and transmitted pulses.
In the boundary value plane wave Fresnel formulation, transient effects are excluded.}

Interestingly, \textcolor{black}{provided the dielectric boundary is sufficiently narrow},
all the \textcolor{black}{Fresnel} plane wave jump conditions are recovered by our 1D QLA simulations [17, 29] except for the ratio of the transmitted to incident electric field amplitudes.  In particular, \textcolor{black}{from the data} we deduced \textcolor{black}{the scaling to more than three digits of accuracy}
\begin{gather*}
\frac{E_{refl}}{E_{inc}	} = \frac{n_1 -n_2}{n_1 + n_2} \qquad ,\quad  \frac{E_{trans}}{E_{inc}	} = \frac{2 n_1}{n_1 + n_2} \sqrt{\frac{n_2}{n_1}} .
\end{gather*}
The QLA simulations lead to an extra factor of $\sqrt{n_2/n_1}$ in the ratio of the transmitted to the incident electric fields. All our simulations
show the ubiquitousness  of this factor, independent of the choices of $n_1$ and $n_2$, as well as various pulse shapes.
Following the simulation results [29], we developed a theory for Gaussian pulses to prove the existence of this factor $\sqrt{n_2/n_1}$.  
\textcolor{black}{In the Appendix, we present new 1D QLA simulation results for an initial rectangular pulse where the Poynting flux can be accurately determined by simple geometry.  Again, we find this ubiquitous extra factor of $\sqrt{n_2/n_1}$.  Moreover it is shown that there is energy conservation during the complete time evolution of the pulse - even during those time instants when the pulse straddles the dielectric boundary layer.   In the Appendix we also show that if we apply the same Fresnel assumptions of reflected and transmitted rectangular pulses, then we do not satisfy energy conservation.  Consequently, it is clear that the Fresnel boundary solution is strictly valid for infinite plane waves.}

Here we are concerned with developing a 2D QLA and performing QLA simulations for the propagation of a 1D electromagnetic pulse in the $x-$direction and incident onto a 2D localized dielectric medium of refractive index $n(x,z)$.  In Sec. 2 we summarize the Khan [28] 8-spinor representation of the inhomogeneous 2D Maxwell equations, while in Sec. 3 we discuss the details of the 2D 16-qubit QLA for $x-z$ dependent fields.  In Sec. 4 we present details of the 2D QLA simulations off various dielectric geometries, showing cases where there is multiple internal reflections within the dielectric medium and conditions under whcih there are no significant internal reflections.  We discuss the $\nabla \cdot \bf{B}$ condition and the role of the collision angle perturbation parameter $\varepsilon$ and its interpretation as a normalized speed of light in the initial medium.  In Sec. 5 we summarize our conclusions.
\textcolor{black}{Also in the Appendix we briefly address how the QLA can be viewed as a quantum algorithm and also from the classical viewpoint
its excellent scaling on classical supercomputers.}
 
\section{Matrix representation of Maxwell equations in a scalar dielectric medium}

The Maxwell equations for wave propagation in a spatially dependent scalar dielectric are,
\begin{align}
\nabla \bsdot \mathbf{D}  \ &= \ \rho_f ,\ & \ \nabla \bsdot \mathbf{B} \ &= \ 0, \label{gauss} \\
\nabla \times \mathbf{E} \ &= \ - \frac{\partial \mathbf{B}}{\partial t},  \ & 
\ \nabla \times \mathbf{H} \ &= \ \mathbf{J}_f + \frac{\partial \mathbf{D}}{\partial t},
\label{fam}
\end{align}
where $\mathbf{E} ( \mathbf{r}, t)$ is the electric, $\mathbf{D}( \mathbf{r}, t) = \epsilon ( \mathbf{r}) \mathbf{E}( \mathbf{r}, t)$ is the displacement
electric field, $\epsilon ( \mathbf{r})$ is the permittivity of the medium, $\mathbf{H}( \mathbf{r}, t) =\mathbf{B}( \mathbf{r}, t) / \mu_0$ is the
magnetic fields, $\mathbf{B}( \mathbf{r}, t)$ is the magnetic flux density, $\mu_0$ is the magnetic permeability of vacuum,
$\rho_f ( \mathbf{r}, t)$ is the free charge density,and $\mathbf{J}_f( \mathbf{r}, t)$ is the free current density.
The electric and magnetic fields are assumed to be real.  
In terms of the Riemann-Silberstein-Weber vectors [28], 
 \begin{equation}
\label{R-S vector}
\mathbf{F^{\pm}} ( \mathbf{r}, t) \ = \  \sqrt{\epsilon ( \mathbf{r})} \ \mathbf{E}( \mathbf{r}, t)  \ \pm \ i \ \frac{\mathbf{B}( \mathbf{r}, t)}{\sqrt{\mu_0}},
\end{equation}
Maxwell equations take on the form,
\begin{equation}
i \frac{\partial \mathbf{F^{\pm}}}{\partial t} = \pm v_{ph} \nabla \times \mathbf{F^{\pm}} 
\pm \frac{1}{2} \nabla v_{ph} \times (\mathbf{F^{\pm}} +
\mathbf{F^{\mp}} )  - i v_{ph} \sqrt{\frac{\mu_0}{2}} 
\mathbf{J}_f,
\label{rswm1}
\end{equation}
\begin{equation}
\nabla \bsdot \mathbf{F^{\pm}} = \frac{1}{2 v_{ph}} \nabla v_{ph} \bsdot (\mathbf{F^{\pm}} +
 \mathbf{F^{\mp}} ) + v_{ph} \sqrt{\frac{\mu_0}{2}} \rho_f .
\label{rswm2}
\end{equation}
where $v_{ph}$ is the phase velocity of the electromagnetic wave,
\begin{equation}
v_{ph} = \frac{1}{\sqrt{\epsilon ( \mathbf{r} ) \mu_0}}   
\end{equation}
From \eqref{rswm1} and \eqref{rswm2}, we note that the coupling between the two field polarizations $\mathbf{F^{\pm}}$ occurs through 
the inhomogeneity in the refractive index $n(\bf{r}) = \sqrt{\epsilon (\bf{r} )}$.

If we assume that there are no free charges or currents, i.e., $\rho_f = 0$ and $\mathbf{J}_f = 0$,
and define an 8-spinor $\Psi^{\pm}$ having the form,
 \begin{equation}
\begin{aligned}
\renewcommand\arraystretch{1.6}
{\Psi^{\pm} }
=  
\begin{pmatrix}
- F^{\pm}_x \pm i F^{\pm}_y    \\
F^{\pm}_z \\
F^{\pm}_z \\
 F^{\pm}_x \pm i F^{\pm}_y  
\end{pmatrix}
,
\end{aligned}
\end{equation}
then, \eqref{rswm1} and \eqref{rswm2} can be written as [28],
\begin{equation}
\begin{aligned}
\renewcommand\arraystretch{1.8}
\frac{\partial}{\partial t} \begin{pmatrix} \Psi^+ \\ \Psi^- \end{pmatrix}
=  
& \renewcommand\arraystretch{2.3}
 - v_{ph}
\begin{pmatrix} \mathbf{M} \bsdot \nabla - \boldsymbol{\Sigma} \bsdot 
\displaystyle{\frac{\nabla \epsilon}{4 \epsilon}}
& i M_z \boldsymbol{\Sigma} \bsdot \displaystyle{\frac{\nabla \epsilon}{2 \epsilon}} \alpha_y \\
i M_z \boldsymbol{\Sigma}^{*} \bsdot \displaystyle{\frac{\nabla \epsilon}{2 \epsilon}} \alpha_y &
\mathbf{M}^{*} \bsdot \nabla - \boldsymbol{\Sigma}^{*} \bsdot 
\displaystyle{\frac{\nabla \epsilon}{4 \epsilon}} \end{pmatrix}
\begin{pmatrix} \Psi^+ \\ \Psi^- \end{pmatrix} .  
\end{aligned}
\label{eightpsi}
\end{equation}
The $\mathbf{M}$ matrices are the tensor products 
of the Pauli spin matrices,
$\boldsymbol{\sigma} = (\sigma_x , \sigma_y , \sigma_z). $
with the $2 \times 2$ identity matrix $\mathbf{I_2}$,
\begin{equation}
\mathbf{M} = \boldsymbol{\sigma} \otimes \mathbf{I_2}  ,  
\ \ \  \Longrightarrow \ \ \ M_z = \sigma_z \otimes \mathbf{I_2}.
\end{equation}
where
\begin{equation}
\label{Pauli_spin}
\sigma_x = 
\begin{pmatrix} 
0 & 1  \\
1 & 0 \\
\end{pmatrix}
\ \ \ 
, \quad \sigma_y = 
\begin{pmatrix} 
0 & - i  \\
i & 0 \\
\end{pmatrix}
\ \ \ 
, \quad \sigma_z = 
\begin{pmatrix} 
1 & 0  \\
0 & -1 \\
\end{pmatrix}.
\end{equation}
The $4 \times 4$ matrices $\boldsymbol{\alpha}$ and $\boldsymbol{\Sigma}$
are defined by,
\begin{equation}
\boldsymbol{\alpha} = 
\begin{pmatrix} 
0 & \boldsymbol{\sigma}  \\
\boldsymbol{\sigma} & 0 \\
\end{pmatrix},
\ \ \ 
\boldsymbol{\Sigma} = 
\begin{pmatrix} 
\boldsymbol{\sigma} & 0  \\
0 & \boldsymbol{\sigma} \\
\end{pmatrix}.
\end{equation}
If we assume a 2D refractive index $n = n(x, z)$, the
8-spinor evolution equation \eqref{eightpsi} reduces to,
\begin{equation}
\label{R_S InHomox1}
\frac{\partial}{\partial t}
\begin{bmatrix}
q_0   \\
q_1  \\
q_2   \\
q_3   \\
\end{bmatrix}  	
= - \frac{1}{n(x,z)} 
\left(
\frac{\partial}{\partial x}
\begin{bmatrix}
q_2  \\
q_3  \\
q_0  \\
q_1  \\  
\end{bmatrix}
+
\frac{\partial}{\partial z}
\begin{bmatrix}
q_0  \\
q_1  \\
-q_2  \\
-q_3  \\ 
\end{bmatrix}
\right)
+ \frac{\partial n / \partial x}{2n^2(x,z)}
\begin{bmatrix}
q_1 + q_6  \\
q_0 - q_7  \\
q_3 - q_4  \\
q_2 + q_5  \\  
\end{bmatrix}
+ \frac{\partial n / \partial z}{2n^2(x,z)}
\begin{bmatrix}
q_0 - q_7  \\
-q_1 - q_6  \\
q_2 + q_5  \\
-q_3 + q_4  \\  
\end{bmatrix}
\end{equation}
\begin{equation}
\label{R_S InHomox1}
\frac{\partial}{\partial t}
\begin{bmatrix}
q_4   \\
q_5  \\
q_6   \\
q_7   \\
\end{bmatrix}  	
= - \frac{1}{n(x,z)} 
\left(
\frac{\partial}{\partial x}
\begin{bmatrix}
q_6  \\
q_7  \\
q_4  \\
q_5  \\  
\end{bmatrix}
+
\frac{\partial}{\partial z}
\begin{bmatrix}
q_4  \\
q_5  \\
-q_6  \\
-q_7  \\ 
\end{bmatrix}
\right)
+ \frac{\partial n / \partial x}{2n^2(x,z)}
\begin{bmatrix}
q_5 + q_2  \\
q_4 - q_3  \\
q_7 - q_0  \\
q_6 + q_1  \\  
\end{bmatrix}
+ \frac{\partial n / \partial z}{2n^2(x,z)}
\begin{bmatrix}
q_4 - q_3  \\
-q_5 - q_2  \\
q_6 + q_1  \\
-q_7 + q_0  \\  
\end{bmatrix}.
\end{equation}
where, in terms of qubits $q_i$ ($i = 0, 1, \ldots 7$),
$\Psi^+ = \left( q_0, q_1, q_2, q_3 \right)^T$, $\Psi^- = \left( q_4, q_5, q_6, q_7 \right)^T$, with the superscript $T$ indicating
the transpose of the row matrix.

\section{Qubit Lattice Algorithm (QLA) for 2D Maxwell Equations}

\subsection{Overview of QLA}

It is convenient to use the Cartesian coordinates to construct the QLA for Maxwell equations since
a 2D QLA reduces to the tensor product of two orthogonal 1D QLAs.
For the 2D refractive index $n(x, z)$, we will assume that the electromagnetic fields are functions of $(x, z, t)$.

Now the 1D QLA is an interleaved sequence of non-commuting collision-stream operators.  The collision operator $C$
entangles the qubits at each lattice site, while the streaming operator $S$ spreads this entanglement throughout the lattice.
Then the basic structure of the evolution operator is,
\begin{equation}
\begin{aligned}
\label{U_X}
U &= \overline{S}_{-} C  \overline{S}_{+} C^{\dagger} \bsdot \: S_{+} C  S_{-} C^{\dagger}, \\
 \overline{U} &= S_{+} C^{\dagger}  S_{-} C \bsdot \: \overline{S}_{-} C^{\dagger}  \overline{S}_{+} C
\end{aligned}
\end{equation}
acting on a qubit vector $Q = \left( q_0, q_1, \ldots \right)^T$.  In \eqref{U_X},
$C^{\dagger}$ is the adjoint of $C$,  $S_{+}$ ($S_-$) streams half of the qubits in $Q$ in the positive (negative) Cartesian direction while
leaving the other half of the qubits untouched,  and $\overline{S}_{+}$ ($\overline{S}_-$)  streams those untouched qubits
in the positive (negative) direction. 
A small parameter $\varepsilon$ is introduced into the unitary collision angle matrix. (We will discuss the significance of
$\varepsilon$ in Section 4.4,  following the simulation results).  The collision and streaming operators are chosen in such a way that when,
\begin{equation}
Q(t+\delta t) = \overline{U} U Q(t),
\end{equation}
is expanded to second order in $\varepsilon$ we recover the 1D parts of Eqs. (12) and (13) that contain the time and spatial derivative of $Q$.
The remaining terms in Eqs. (12) and (13), which depend on the derivative of the refractive index are recovered through a potential operator in
which the collision operator depends on $\varepsilon$. 

Interestingly, the 2D QLA can use the same dimensionality of the qubit vector $Q$ as in 1D but with $Q=Q(x,z,t)$.
A simple notational generalization of Eq. (15) leads to,
\begin{equation}
Q(t+\delta t) = \overline{U}_Z U_Z . \overline{U}_X U_X Q(t)
\end{equation}
For appropriately chosen collision-stream operators we recover the required
$\partial / \partial t$, and the spatial derivative terms $\partial / \partial x$ and $\partial / \partial z$ in the limit of
long times and long spatial scales with $\varepsilon << 1$.

To determine the collision operator for x-dependence, we note in Eqs. (12)-(13) that the qubit
couplings of $\partial / \partial t$ with $\partial / \partial x$ requires the coupling of
$q_0$ - $q_2$, $q_1$ - $q_3$, $q_4$ - $q_6$, $q_5$ - $q_7$.  This can readily be achieved by an $8 \cross 8$ collision matrix.  
However,  to recover the coupling of $\partial / \partial t$ - $\partial / \partial z$ in Eqs. (12)-(13) we note that in the 8-qubit
representation has  diagonal coupling $q_0$ - $q_0$, $q_1$ - $q_1$, $q_2$ - $q_2$, $...$.  Since entanglement requires a collision
operator to act on at least 2 qubits we must extend to a 16-qubit representation.
This problem of diagonal coupling of an 8-qubit system can be traced back to the diagonal structure of the Pauli spin matrix $\sigma_z$.  
Since both $U_X$ and $U_Z$ act on the same qubit vector $Q(t)$, it is reasonable to develop the 2D QLA for a 16-component qubit vector.

\subsection{16-qubit QLA for 1D propagation in the $x$-direction}

A 16-qubit formulation for 1D QLA propagation in the x-direction can be determined from the
unitary $16 \cross 16$ collision matrix $\overline{C}_X$ and the $8 \cross 8$ submatrices $C_{8X}$,
\begin{equation}
\overline{C}_{X} = 
    \begin{bmatrix}
    C_{8X} (\theta)  &  0 \\
    0  & C_{8X} (\theta) 
    \end{bmatrix}
\end{equation}
where
\begin{equation}
C_{8X} (\theta) = 
  \begin{bmatrix}
  \cos \theta & 0 &  0 & 0 & \sin \theta & 0 & 0 & 0   \\
  0  &  \cos \theta  &  0  &  0  &  0  &  \sin \theta  &  0  &  0 \\
 0  &  0  &  \cos \theta  &  0  & 0 & 0 & \sin \theta & 0   \\
  0  &  0  &  0  &  \cos \theta  &  0  &  0  &  0  &  \sin \theta \\ 
 -\sin \theta & 0 & 0 & 0  &  \cos \theta & 0 &  0 & 0 \\
   0 & -\sin \theta & 0 & 0  &  0  &  \cos \theta  &  0  &  0\\
    0 & 0 & -\sin \theta & 0  & 0  &  0  &  \cos \theta  &  0 \\
     0 & 0 & 0 & -\sin \theta &  0  &0 &  0  &  \cos \theta  \\
   \end{bmatrix}.
\end{equation}    
with collision angle
\begin{equation}
\theta  = \frac{\varepsilon}{4 n(x,z)}.
\end{equation}
The streaming operator $\overrightarrow{S}$ will stream the 8-qubit subset $[\overline{q}_0, \overline{q}_1, \overline{q}_2, \overline{q}_3, \overline{q}_8, \overline{q}_9, \overline{q}_{10}, \overline{q}_{11}]$, while the streaming operator $\overline{\overrightarrow{S}}$ will stream the remaining 8-qubit subset  $[\overline{q}_4, \overline{q}_5, \overline{q}_6, \overline{q}_7, \overline{q}_{12}, \overline{q}_{13}, \overline{q}_{14}, \overline{q}_{15}]$.

To complete the 16-qubit 1D QLA for $x$-propagation, we need to define the potential
collision operators that will recover the $\partial n/\partial x$ terms in the (reduced) Eqs. (12)-(13).  There will be two such potential operators.
The first will have block diagonal structure based on the $2 \cross 2$ matrix
\begin{equation}
P_1(\gamma) = 
    \begin{bmatrix}
    \cos\gamma  &  \sin\gamma \\
     \sin\gamma  &  \cos\gamma   
    \end{bmatrix}
\end{equation}
with
\begin{equation}
\overline{P}_{1X} (\gamma) = 
  \begin{bmatrix}
 P_1(\gamma)& 0 &  0 & 0 &0 & 0 & 0 & 0   \\
  0  &  P_1(\gamma) &  0  &  0  &  0  &0  &  0  &  0 \\
 0  &  0  & P_1(\gamma)  &  0  & 0 & 0 & 0 & 0   \\
  0  &  0  &  0  & P_1(\gamma)  &  0  &  0  &  0  &  0 \\ 
 0& 0 & 0 & 0  & P_1(\gamma) & 0 &  0 & 0 \\
   0 & 0 & 0 & 0  &  0  & P_1(\gamma)  &  0  &  0\\
    0 & 0 &0 & 0  & 0  &  0  &  P_1(\gamma)  &  0 \\
     0 & 0 & 0 & 0 &  0  &0 &  0  & P_1(\gamma)  \\
   \end{bmatrix}.
\end{equation} 
The potential collision angle is,
\begin{equation}
\gamma = \varepsilon^2 \frac{•\partial n/ \partial x}{2 n^2}.
\end{equation}
$\overline{P}_{1X} (\gamma) $ is Hermitian but not unitary. 
The second of the potential collision operators, $\overline{P}_{2X}(\gamma)$, is a sparse unitary operator based on two $8 \cross 8$ subpotential matrices,
\begin{equation}
\overline{P}_{2X}(\gamma) = 
    \begin{bmatrix}
    P_{11} (\gamma) & P_{12} (\gamma \\
     P_{12} (\gamma) & P_{11} (\gamma) 
    \end{bmatrix} .
\end{equation}
These $8 \cross 8$ matrices are given by,
\begin{equation}
P_{11} (\gamma) = 
  \begin{bmatrix}
\cos\gamma & 0 &  0 & 0 &0 & 0 & 0 & 0   \\
  0  & \cos\gamma &  0  &  0  &  0  &0  &  0  &  0 \\
 0  &  0  & \cos\gamma  &  0  & 0 & 0 & 0 & 0   \\
  0  &  0  &  0  & \cos\gamma  &  0  &  0  &  0  &  0 \\ 
 0& 0 & 0 & 0  &\cos\gamma& 0 &  0 & 0 \\
   0 & 0 & 0 & 0  &  0  & \cos\gamma &  0  &  0\\
    0 & 0 &0 & 0  & 0  &  0  &  \cos\gamma  &  0 \\
     0 & 0 & 0 & 0 &  0  &0 &  0  & \cos\gamma  \\
   \end{bmatrix}.
\end{equation} 
and
\begin{equation}
P_{12} (\gamma) = 
  \begin{bmatrix}
0 & 0 &  0 & 0 & \sin\gamma & 0 & 0 & 0   \\
  0  & 0&  0  &  0  & 0  & -\sin\gamma   &  0  &  0 \\
 0  &  0  &0  &  0  & 0 & 0 & \sin\gamma  & 0   \\
  0  &  0  &  0  & 0  &  0  &  0  &  0  &  -\sin\gamma  \\ 
 - \sin\gamma& 0 & 0 & 0  &0& 0 &  0 & 0 \\
   0 &  \sin\gamma & 0 & 0  &  0  & 0&  0  &  0\\
    0 & 0 &- \sin\gamma & 0  & 0  &  0  & 0 &  0 \\
     0 & 0 & 0 &  \sin\gamma &  0  &0 &  0  & 0 \\
   \end{bmatrix}.
\end{equation} 

\subsection{16-qubit QLA for 1D z-propagation}

For the 16-qubit vector $\overline{Q}$ an appropriate unitary collision operator which couples the qubits $\overline{q}_0 - \overline{q}_2$,  $\overline{q}_1 - \overline{q}_3$,   $\overline{q}_4 - \overline{q}_{6}$, $\overline{q}_5 - \overline{q}_7$,  $\overline{q}_8 - \overline{q}_{10}$,  $\overline{q}_9 - \overline{q}_{11}$,  $\overline{q}_{12} - \overline{q}_{14}$, $\overline{q}_{13} - \overline{q}_{15}$
is derived from the $4 \cross 4$ block diagonal structure,
\begin{equation}
\label{C4x4}
\overline{C}_Z (\theta) = 
\begin{bmatrix}
C_4 (\theta) & 0 & 0 & 0   \\
0  &  C_4 (\theta)^{T}  & 0 & 0  \\  
0  & 0 & C_4 (\theta)  &  0 \\
 0  &  0 & 0 & C_4 (\theta)^{T} \\ 
\end{bmatrix},
\end{equation}
with the $4 \cross 4$ submatrix,
\begin{equation}
\label{Csub4x4}
C_4 (\theta) = 
\begin{bmatrix}
\cos \theta & 0 & \sin \theta & 0   \\
0  &  \cos \theta  & 0 & \sin \theta  \\  
-\sin \theta  & 0 &  \cos \theta  &  0 \\
 0  &  -\sin \theta & 0 & \cos \theta  \\ 
\end{bmatrix},
\end{equation} 
where $\theta$ given by Eq. (19).

The streaming operator $\overrightarrow{S}_Z$ will stream the 8-qubit subset $[\overline{q}_0, \overline{q}_1, \overline{q}_4, \overline{q}_5, \overline{q}_8, \overline{q}_9, \overline{q}_{12}, \overline{q}_{13}]$, while the streaming operator $\overline{\overrightarrow{S}}_Z$ will stream the remaining 8-qubit subset  $[\overline{q}_2, \overline{q}_3, \overline{q}_6, \overline{q}_7, \overline{q}_{10}, \overline{q}_{11}, \overline{q}_{14}, \overline{q}_{15}]$.

To recover the $\partial n / \partial z$ terms in the (reduced) Eqs. (12) - (13), we introduce two potential collision operators.
The first of the potential collision operators has a form that mimics the coupling of the unitary collision matrix $\overline{C}_Z$ with its $4 \cross 4$ diagonal block structure,
\begin{equation}
\label{C4x4}
\overline{P}_{1Z} (\gamma') = 
\begin{bmatrix}
P_4 (\gamma') & 0 & 0 & 0   \\
0  &  P_4 (\gamma') & 0 & 0  \\  
0  & 0 & P_4 (\gamma') &  0 \\
 0  &  0 & 0 & P_4 (\gamma') \\ 
\end{bmatrix},
\end{equation} 
where
\begin{equation}
\label{Csub4x4}
P_4 (\gamma') = 
\begin{bmatrix}
\cos \gamma' & 0 &- \sin \gamma' & 0   \\
0  &  \cos \gamma'  & 0 & - \sin \gamma' \\  
- \sin \gamma' & 0 & \cos \gamma'  &  0 \\
 0  &  - \sin \gamma' & 0 &\cos \gamma' \\ 
\end{bmatrix}.
\end{equation}
This evolution potential collision operator is Hermitian.
The second  potential collision operator has diagonal-like structure of two $8 \cross 8$ matrices,
\begin{equation}
\label{Csub4x4}
\overline{P}_{2Z} (\gamma') = 
\begin{bmatrix}
P_{81} (\gamma') & P_{82} (\gamma')   \\
P_{82} (\gamma') & P_{81} (\gamma')  \\  
\end{bmatrix},
\end{equation}
where
\begin{equation}
\label{PV81 8x8}
P_{81} (\gamma') = 
  \begin{bmatrix}
  \cos \gamma' & 0 & 0 & 0 & 0 & 0 & 0 & 0   \\
  0  &  \cos \gamma'  &  0  & 0  &  0  &  0  &  0  &  0 \\
  0  &  0  &  \cos \gamma'  &  0  & 0 & 0 & 0 & 0   \\
  0  &  0  &  0  &  \cos \gamma'  &  0  &  0  &  0  &  0 \\ 
   0 & 0 & 0 & 0  &  \cos \gamma' & 0 &  0 & 0 \\
   0 & 0 & 0 & 0  &  0  &  \cos \gamma' &  0  &  0 \\
    0 & 0 & 0 & 0  & 0  &  0  &  \cos \gamma'  &  0 \\
     0 & 0 & 0 & 0  &  0  & 0  &  0  &  \cos\gamma'  \\
   \end{bmatrix},
\end{equation} 
and,
\begin{equation}
\label{PV82 8x8}
P_{82} (\gamma') = 
  \begin{bmatrix}
 0 & 0 & 0 & 0 & 0 & 0 & 0 & - \sin \gamma'   \\
  0  & 0  &  0  & 0  &  0  &  0  &  - \sin \gamma'  &  0 \\
  0  &  0  & 0 &  0  & 0 & - \sin \gamma' & 0 & 0   \\
  0  &  0  &  0  & 0 &  - \sin \gamma'  &  0  &  0  &  0 \\ 
   0 & 0 & 0 &  \sin \gamma'   &  0 & 0 &  0 & 0 \\
   0 & 0 &  \sin \gamma'  & 0  &  0  &  0 &  0  &  0 \\
    0 &  \sin \gamma'  & 0 & 0  & 0  &  0  &  0  &  0 \\
      \sin \gamma'  & 0 & 0 & 0  &  0  & 0  &  0  &  0  \\
   \end{bmatrix}.
\end{equation} 
The potential collision angle is similar to that for $x$-propagation except that the derivative on the refractive index is with respect to $z$,
\begin{equation}
\gamma' = \varepsilon^2 \frac{•\partial n/ \partial z}{2 n^2}.
\end{equation}

\subsection{16-qubit QLA for 2D for propagation in the $x-z$ plane}

Piecing together the 1D propagation in $x$ and $z$ directions,
we obtain the 2D QLA acting on the 16-qubit vector. Thus, for the $x$-dependence we have the unitary collide-stream sequence in the $x$-direction,
\begin{equation}
U_X = \overline{S}_{-X} \overline{C}_X \overline{S}_{+X} \overline{C}_X^{\dagger}.S_{+X} \overline{C}_X S_{-X}\overline{C}_X^{\dagger}
\end{equation}
\begin{equation}
\overline{U}_X = S_{+X} \overline{C}_X^{\dagger} S_{-X}\overline{C}_X . \overline{S}_{-X} \overline{C}_X^{\dagger} \overline{S}_{+X} \overline{C}_X.
\end{equation}
For the $z$-dependence, the corresponding unitary collide-stream structure is,
\begin{equation}
U_Z = \overline{S}_{-Z} \overline{C}_Z \overline{S}_{+Z} \overline{C}_Z^{\dagger}.S_{+Z} \overline{C}_Z S_{-Z}\overline{C}_Z^{\dagger}
\end{equation}
\begin{equation}
\overline{U}_Z = S_{+Z} \overline{C}_Z^{\dagger} S_{-Z}\overline{C}_Z . \overline{S}_{-Z} \overline{C}_Z^{\dagger} \overline{S}_{+Z} \overline{C}_Z..
\end{equation}

The 2D 16-qubit QLA algorithm is,
\begin{equation}
\overline{Q}(t+\delta t) = \overline{P}_{1Z} \overline{P}_{2Z} . \overline{U}_Z. U_Z.\overline{P}_{1X} \overline{P}_{2X} . \overline{U}_X. U_X \overline{Q}(t)
\end{equation}
In the long time, long spatial scale limit ($\varepsilon << 1$)
we obtain a set of 16 partial differential equations for $\big [\overline{q_0} , ...,\overline{q_{15}} \big ]$.
We recover the 8-spinor 2D representation, $\big [q_0 , ... , q_7 \big ]$, of Maxwell equations to $O(\varepsilon^2)$, Eq. (12)-(13), by the identification,
\begin{equation}
q_0 = \overline{q}_0+\overline{q}_2  ,\quad  q_1 = \overline{q}_1+\overline{q}_3,\quad q_2 = \overline{q}_4+\overline{q}_6,\quad q_3 = \overline{q}_5+\overline{q}_7,
\end{equation}
\begin{equation}
q_4 = \overline{q}_8+\overline{q}_{10}  , \quad q_5 = \overline{q}_9+\overline{q}_{11},\quad q_6 = \overline{q}_{12}+\overline{q}_{14},\quad q_7 = \overline{q}_{13}+\overline{q}_{15},
\end{equation}

The initial conditions correspond to the electromagnetic field components of the pulse at $t = 0$.
Using the Riemann-Silberstein-Weber variables, we obtain the initial condition on the 8-spinor $(q_0 , .... q_7)$.
For the 16-qubit $\overline{Q}(t=0)$  the initial condition is,

\begin{equation}
 \overline{q}_0 = \overline{q}_2 = q_0/2  , \quad \overline{q}_1=\overline{q}_3=q_1/2 , \quad \overline{q}_4=\overline{q}_6 = q_2/2 ,\quad \overline{q}_5=\overline{q}_7 =q_3/2,
\end{equation}
\begin{equation}
 \overline{q}_8=\overline{q}_{10} = q_4/2  ,\quad \overline{q}_9=\overline{q}_{11} =q_5/2,\quad \overline{q}_{12}=\overline{q}_{14} = q_6/2,\quad \overline{q}_{13}=\overline{q}_{15} = q_7/2 .
\end{equation}

\section{2D QLA SIMULATIONS}
QLA simulations will be presented for an initial 1D  Gaussian pulse propagating in the $x-$direction that interacts with
a small 2D cylindrical dielectric with refractive index $n(x,z)$ (section 4.1), and with a small 2D conic dielectric (section 4.2).

\subsection{Gaussian Pulse interacting with a 2D Dielectric Cylinder  :  $n_1 < n_2$ }
Consider a 1D Gaussian electromagnetic pulse propagating in the $x-$direction with polarization $E_y(x,t=0)$ and with $B_z(x,t=0)$ in a vacuum, $n_1 = 1$.  It propagates towards a dielectric cylinder located in the center of square grid $l \cross L$, with peak refractive index $n=3$ (see Fig. 1.)
\begin{figure}[h] \ 
\begin{center}
\vspace{0cm} 
\includegraphics[width=3.2in]{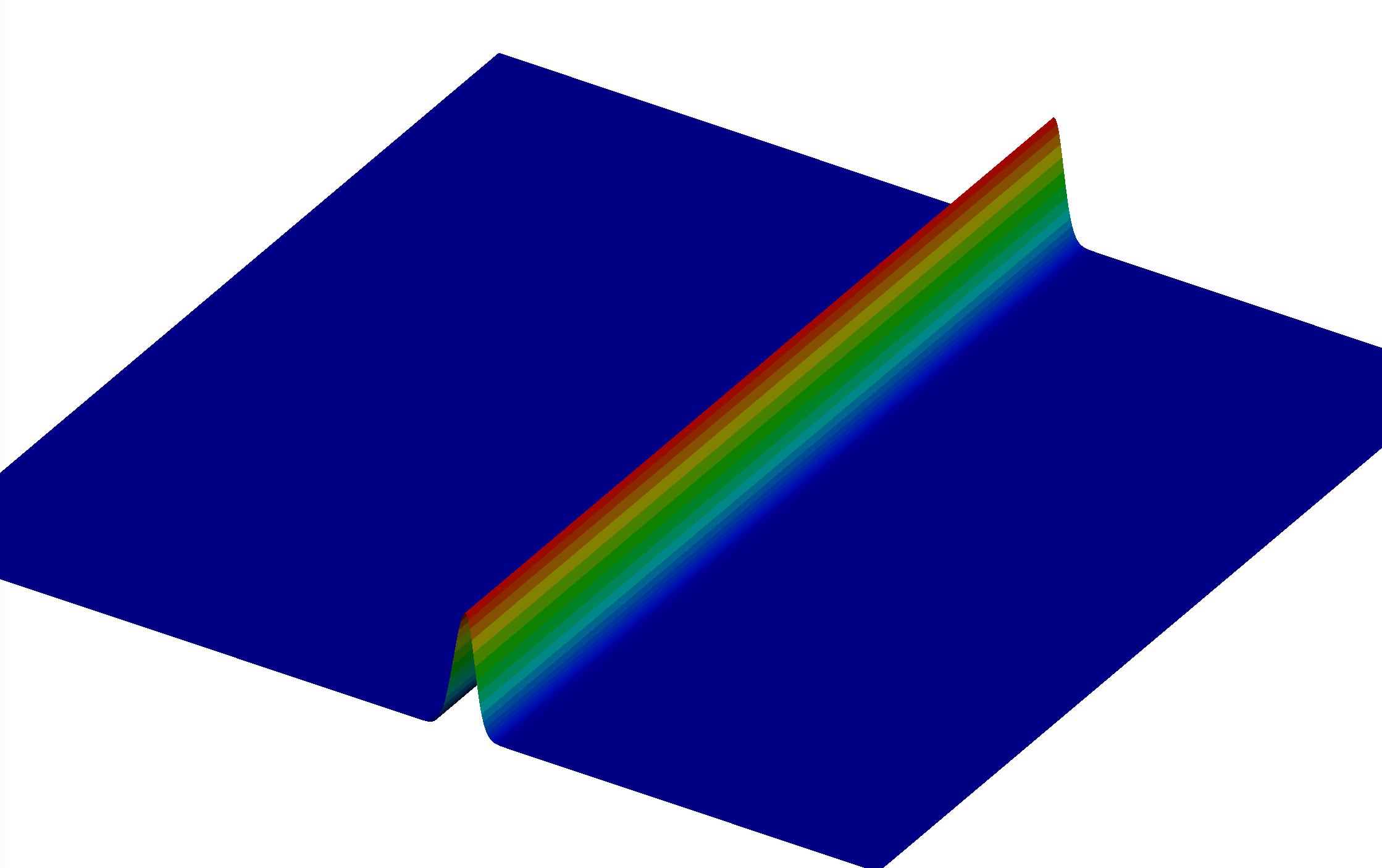}
\includegraphics[width=3.2in]{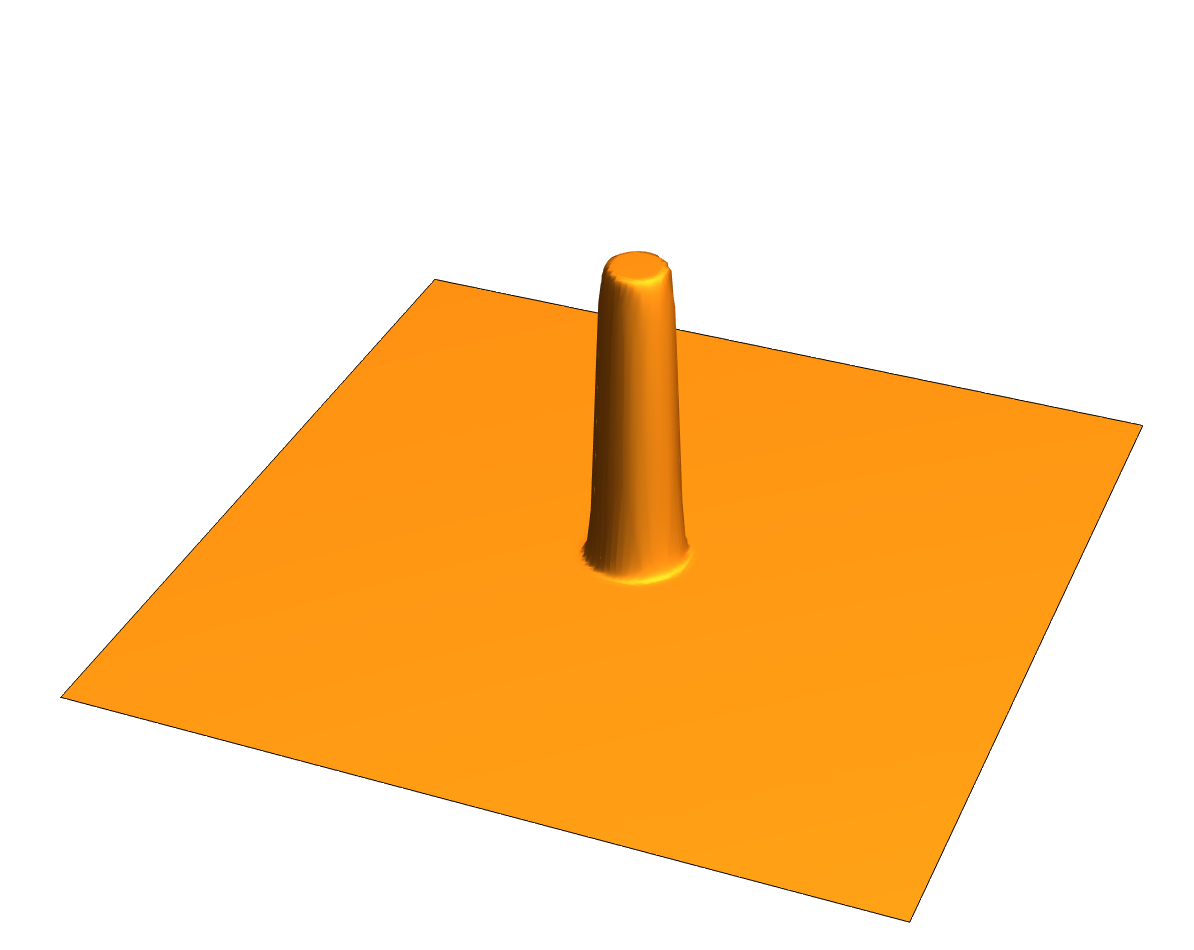}
(a)  $E_y$ initial field , \quad  \quad  \quad \qquad  (b)  dielectric cylinder
\caption{(a) The $E_y$ field of the initial electromagnetic Gaussian pulse as it propagates in the vacuum ($n_1 = 1$), (b) a localized dielectric cylinder with peak refractive index $n(x=L/2,z=L/2) =3$.  Grid $L^2 = 8192^2$.
}
\end{center}
\vspace{0cm} 
\end{figure} 

When the 1D pulse starts to interact with the 2D dielectric cylinder, the electromagnetic fields take on 2D structure.  In Fig. 2 we plot the electric field component $E_y(x,z)$, both from the perspective of (a) $E_y > 0$, and (b) $E_y < 0$ 
\begin{figure}[!] \ 
\begin{center}
\vspace{0cm} 
\includegraphics[width=3.2in]{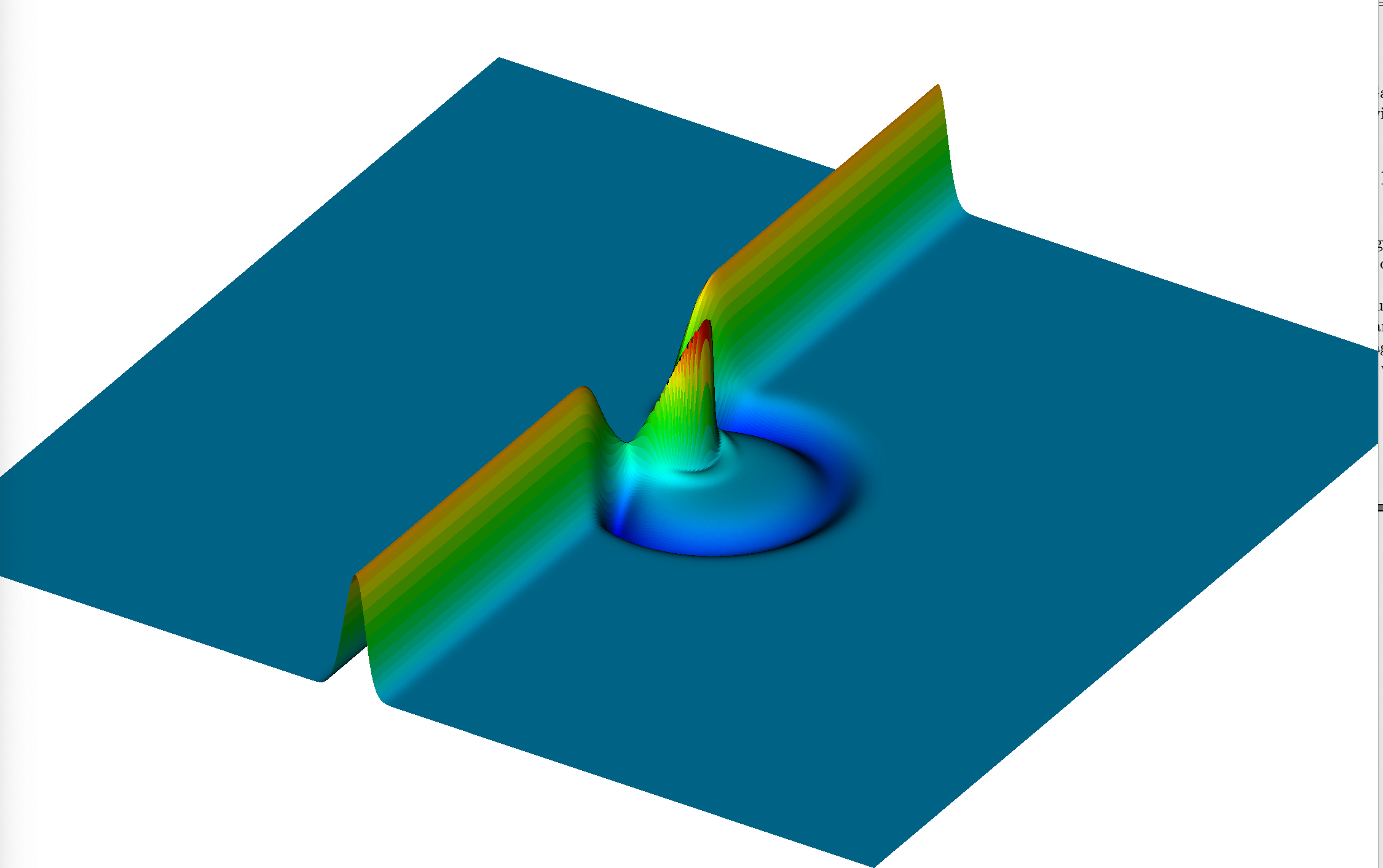}
\includegraphics[width=3.2in]{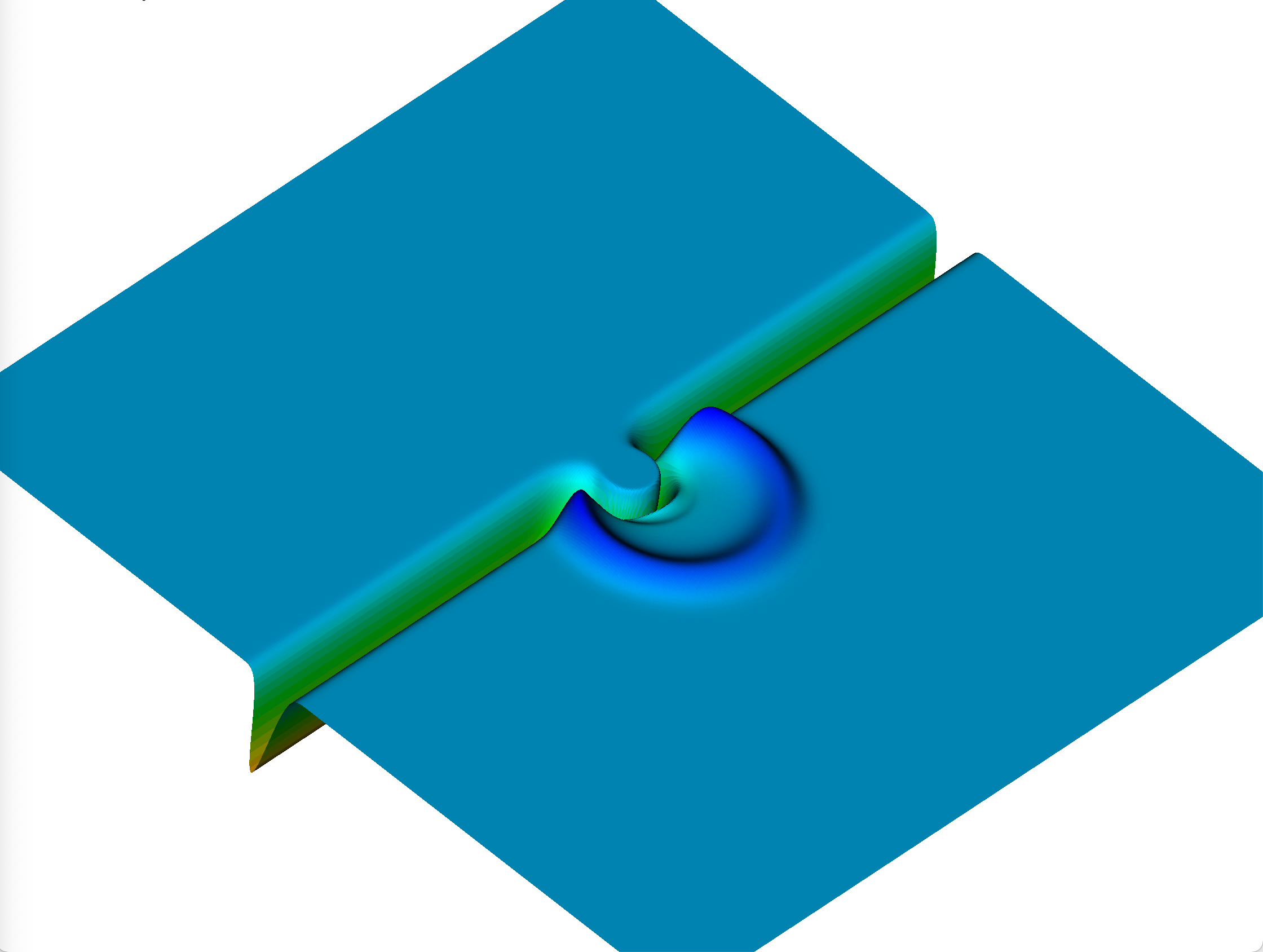}
(a)  $E_y > 0$ perspective , \quad  \quad  \quad \qquad  (b)  $E_y < 0$ perspective
\caption{(a) The $E_y > 0$ field after the pulse has started to overlap into the dielectric cylinder,  (b) the corresponding $E_y$ field, but from the perspective $E_y < 0$.  There is a phase change of $\pi$ in the reflected $E_y$ ring since $n_2 > n_1$.
}
\end{center}
\vspace{0cm} 
\end{figure} 
That part of the pulse which is propagating within the dielectric cylinder is moving with a slower phase velocity than that part of the pulse that remains in the vacuum and hence will lag the pulse that is in $n_1$ region.  
More interesting is the ring region of the reflected pulse in which $E_y < 0$.  This is readily seen in Fig. 2(b) and is somewhat expected from analogy with the 1D scattering of a pulse propagating from region $n_1$ to region $n_2$ with $n_2 > n_1$.

As the main part of the pulse keeps propagating in the x-direction, there are now internal reflections$/$transmissions of the fields from the dielectric cylinder since the boundary layer region between the two media is quite thin.  These give rise to interesting wavefronts emanating from the dielectric cylinder, Fig. 3
\begin{figure}[!] \ 
\begin{center}
\vspace{0cm} 
\includegraphics[width=3.2in]{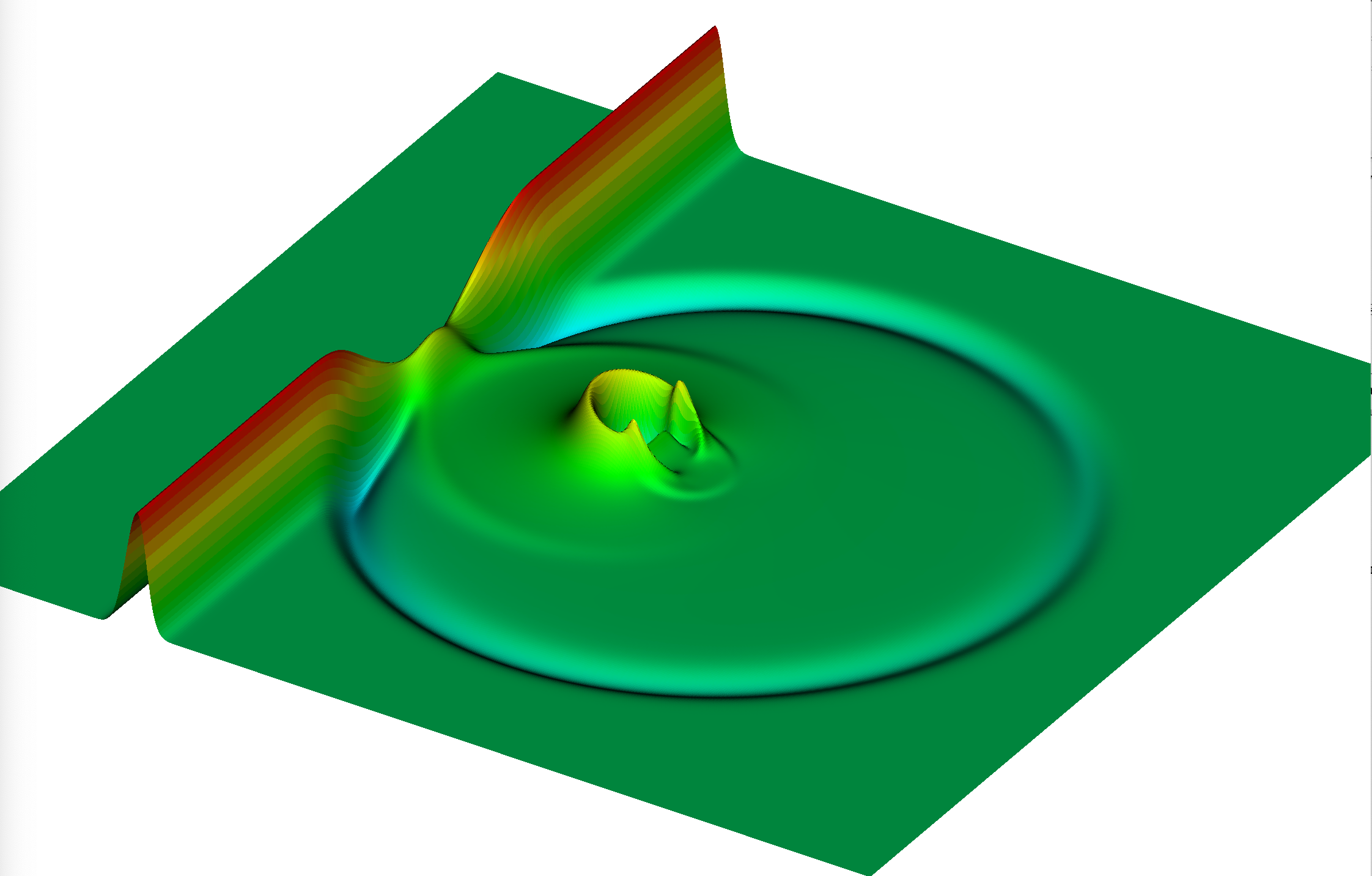}
\includegraphics[width=3.2in]{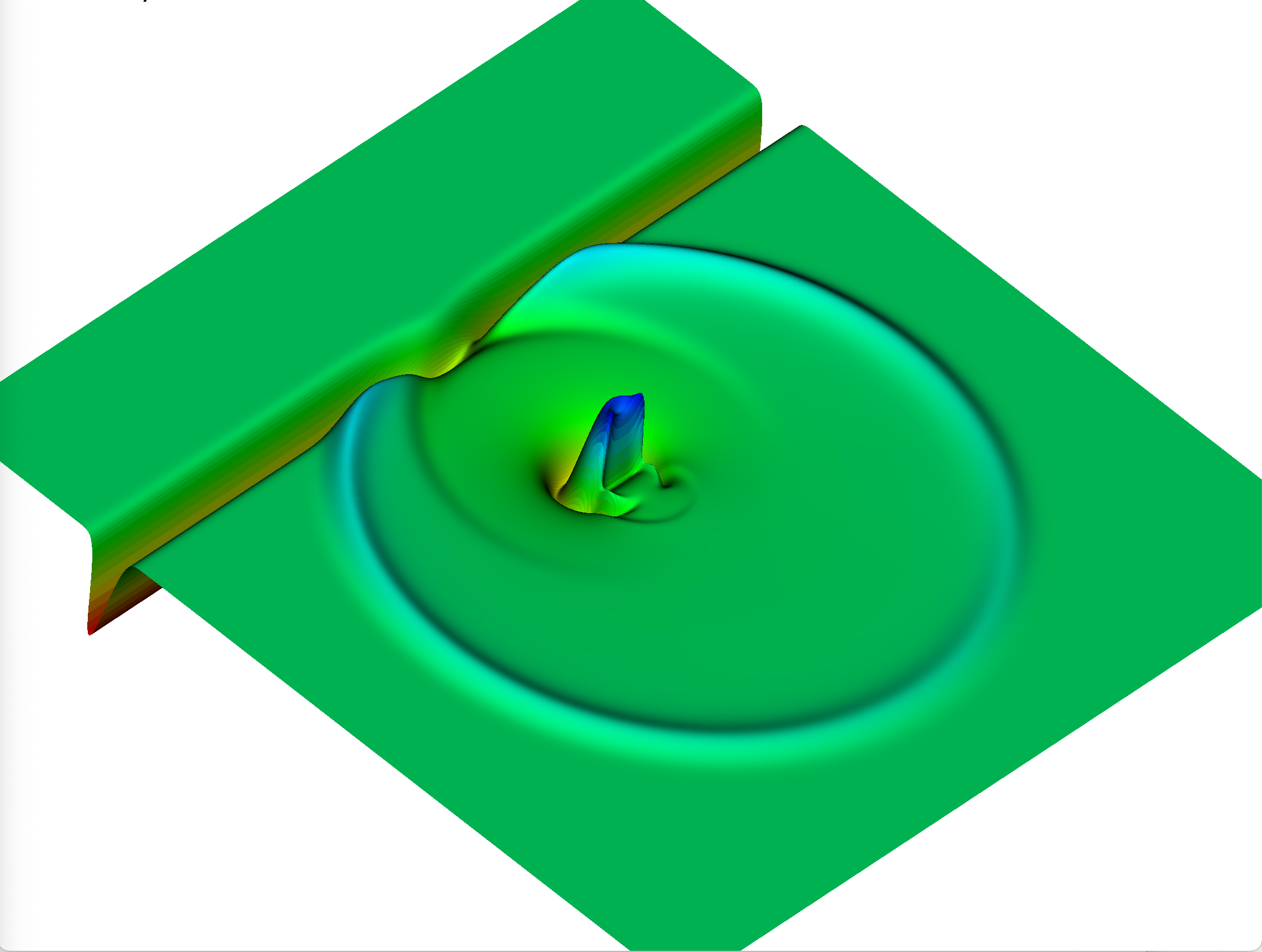}
(a)  $E_y > 0$ perspective , \quad  \quad  \quad \qquad  (b)  $E_y < 0$ perspective
\caption{(a) The $E_y > 0$ field after parts of the pulse have undergone reflection$/$transmission within the dielectric cylinder,  (b) the corresponding $E_y$ field, but from the perspective $E_y < 0$.  The initial ring of $E_y < 0$ in Fig. 2 is radiated away from the cylinder while new wavefronts are formed from the transient fields trapped within the dielectric.
}
\end{center}
\vspace{0cm} 
\end{figure} 

\begin{figure}[!] \ 
\begin{center}
\vspace{0cm} 
\includegraphics[width=3.2in]{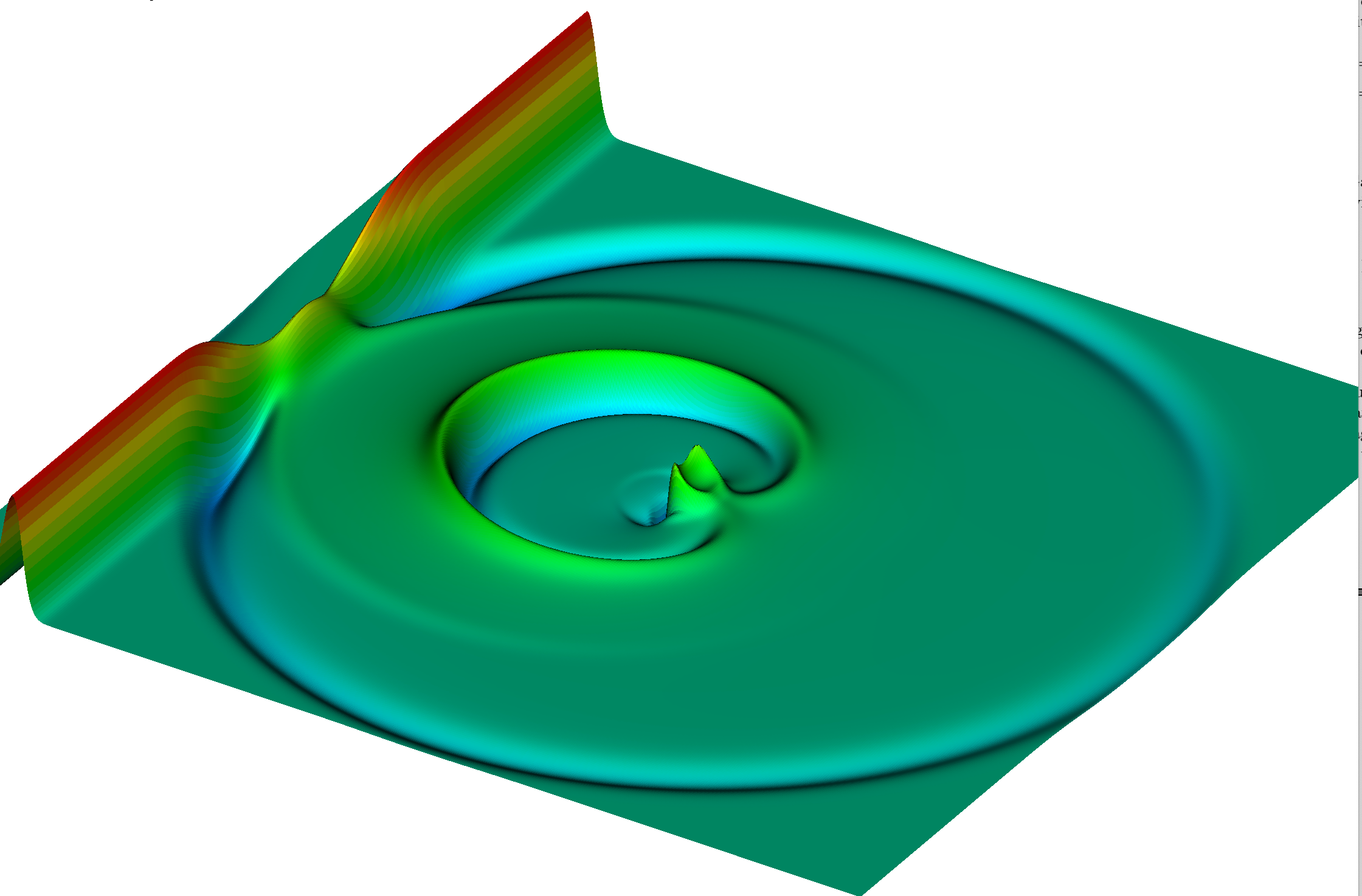}
\includegraphics[width=3.2in]{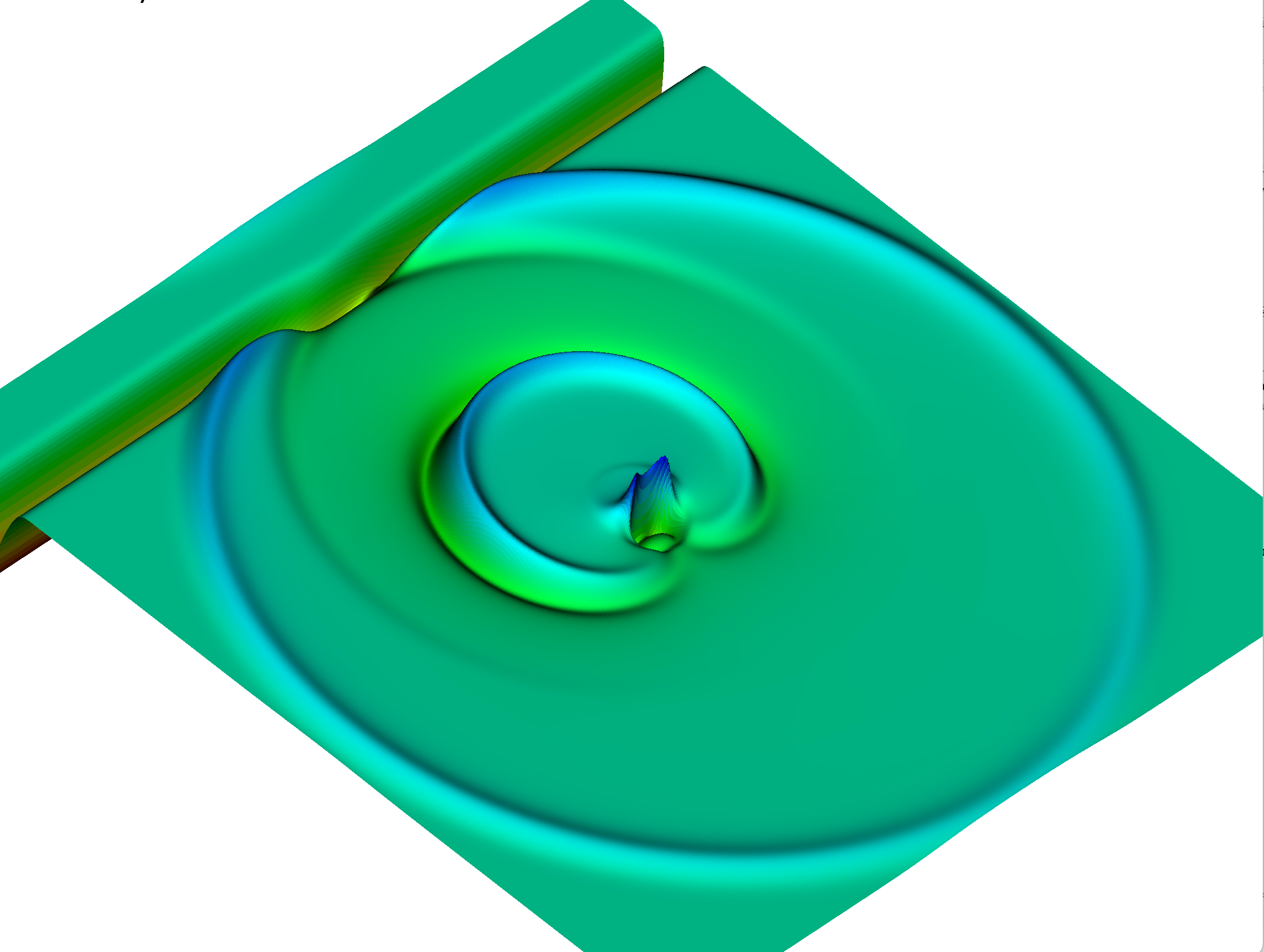}
(a)  $E_y > 0$ perspective , \quad  \quad  \quad \qquad  (b)  $E_y < 0$ perspective
\caption{The $E_y$ field later in the initial value simulation from the perspective of (a) The $E_y > 0$, and (b) $E_y < 0$.  
}
\end{center}
\vspace{0cm} 
\end{figure} 
In Fig. 4 we plot the $E_y$ near the end of the QLA simulation.

The QLA is modeling the time evolution of the full Maxwell equations, including the two divergence equations that are typically treated as initial conditions.  To ensure that $\nabla \cdot \mathbf{B} = 0$ a $B_x(x,z)$ field component must be generated when the 1D pulse interacts with the 2D dielectric cylinder with $n(x,z)$.   In Figs. 5-7  we plot the $B_x$-field at the time instances corresponding to the time snapshots  of Fig. 2-4.

\begin{figure}[!] \ 
\begin{center}
\vspace{0cm} 
\includegraphics[width=3.2in]{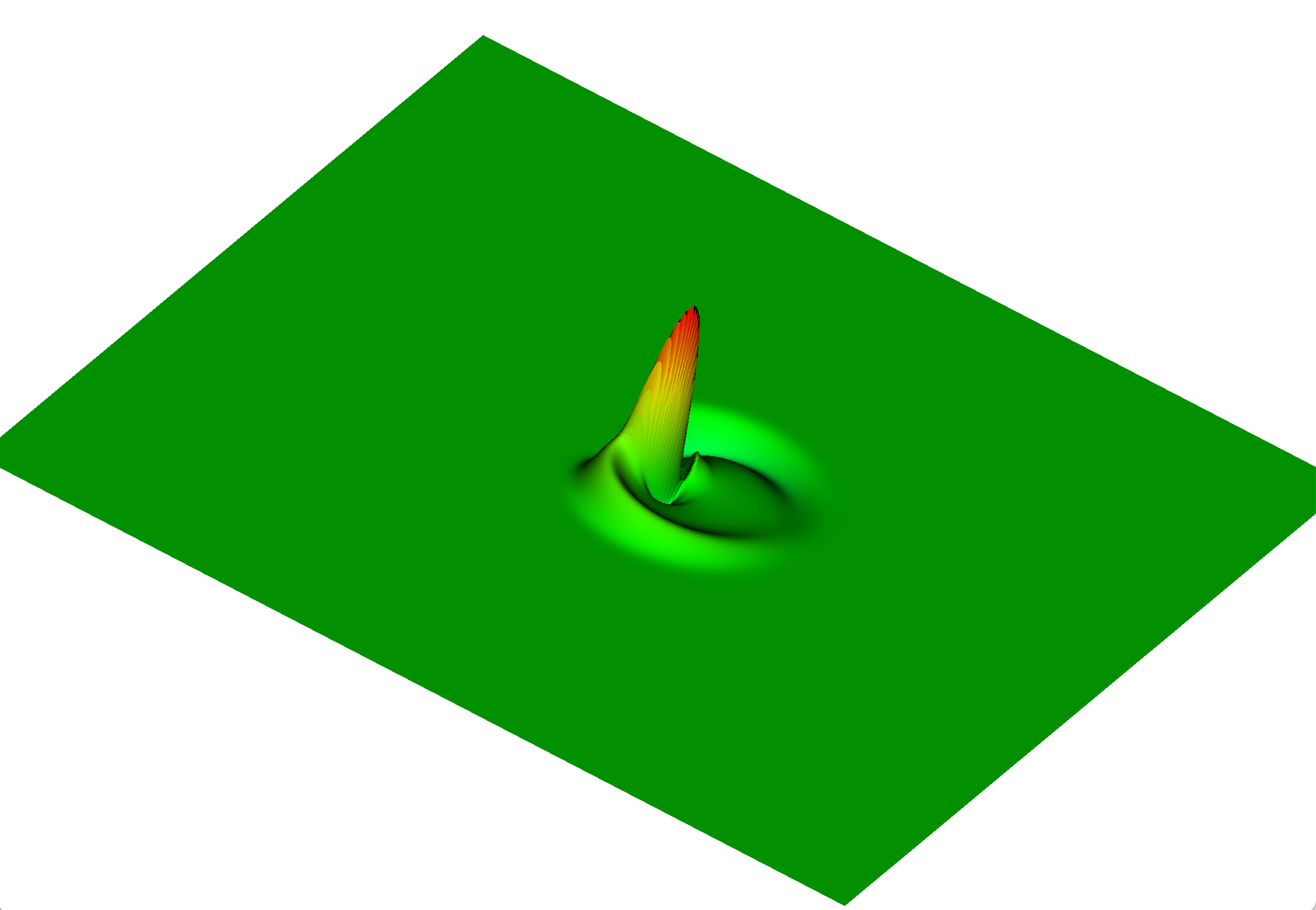}
\includegraphics[width=3.2in]{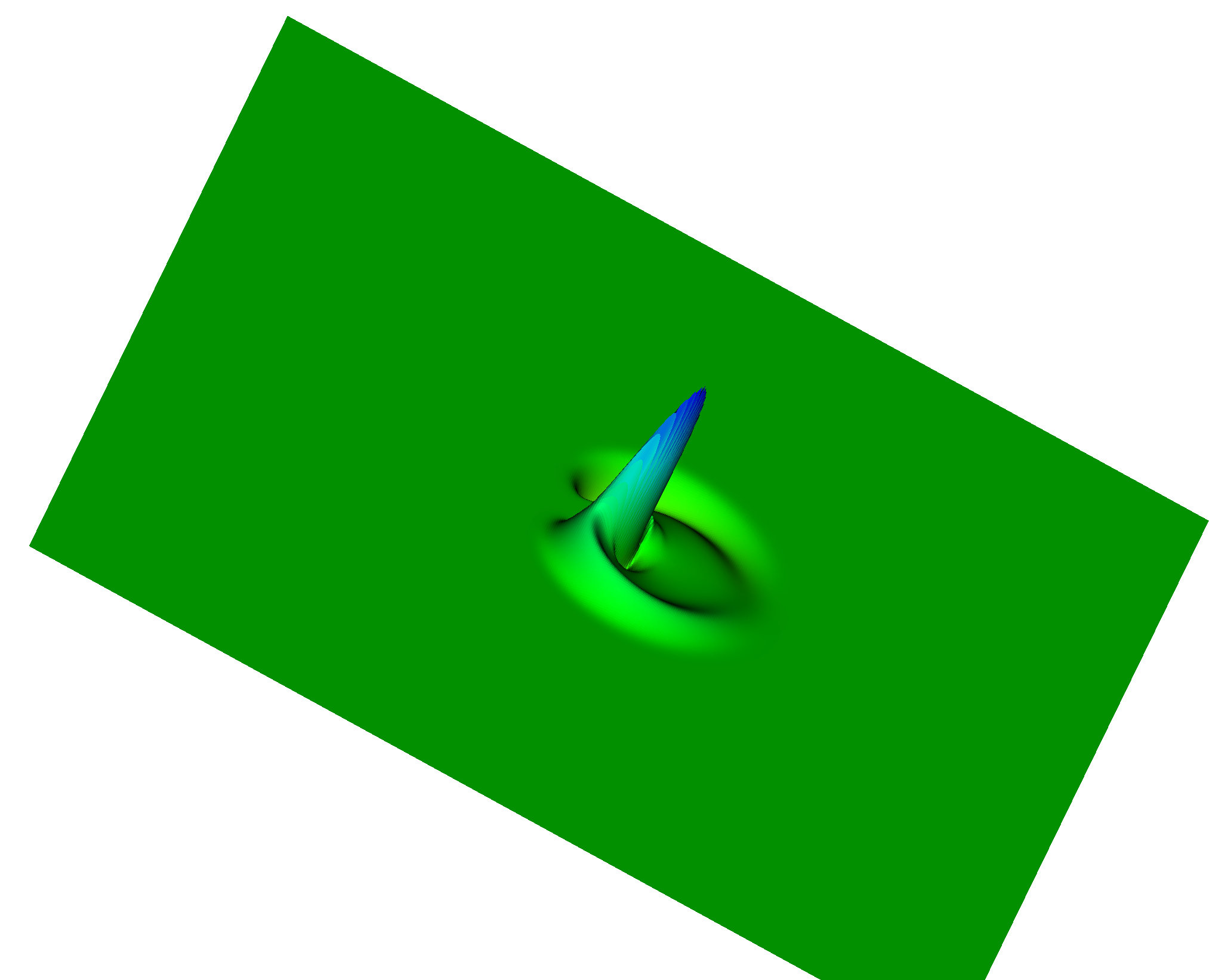}
(a)  $B_x > 0$ perspective , \quad  \quad  \quad  (b)  $B_x < 0$ perspective - c.f.  Fig. 2
\caption{The $B_x(x,z)$ field generated from the interaction of the 1D pulse with the 2D dielectric.  Time snapshot corresponds to that of Fig. 2 and is on the same scale as Fig. 2.  (a) The $B_x > 0$, and (b) $B_x < 0$.  
}
\end{center}
\vspace{0cm} 
\end{figure} 

\begin{figure}[!] \ 
\begin{center}
\vspace{0cm} 
\includegraphics[width=3.2in]{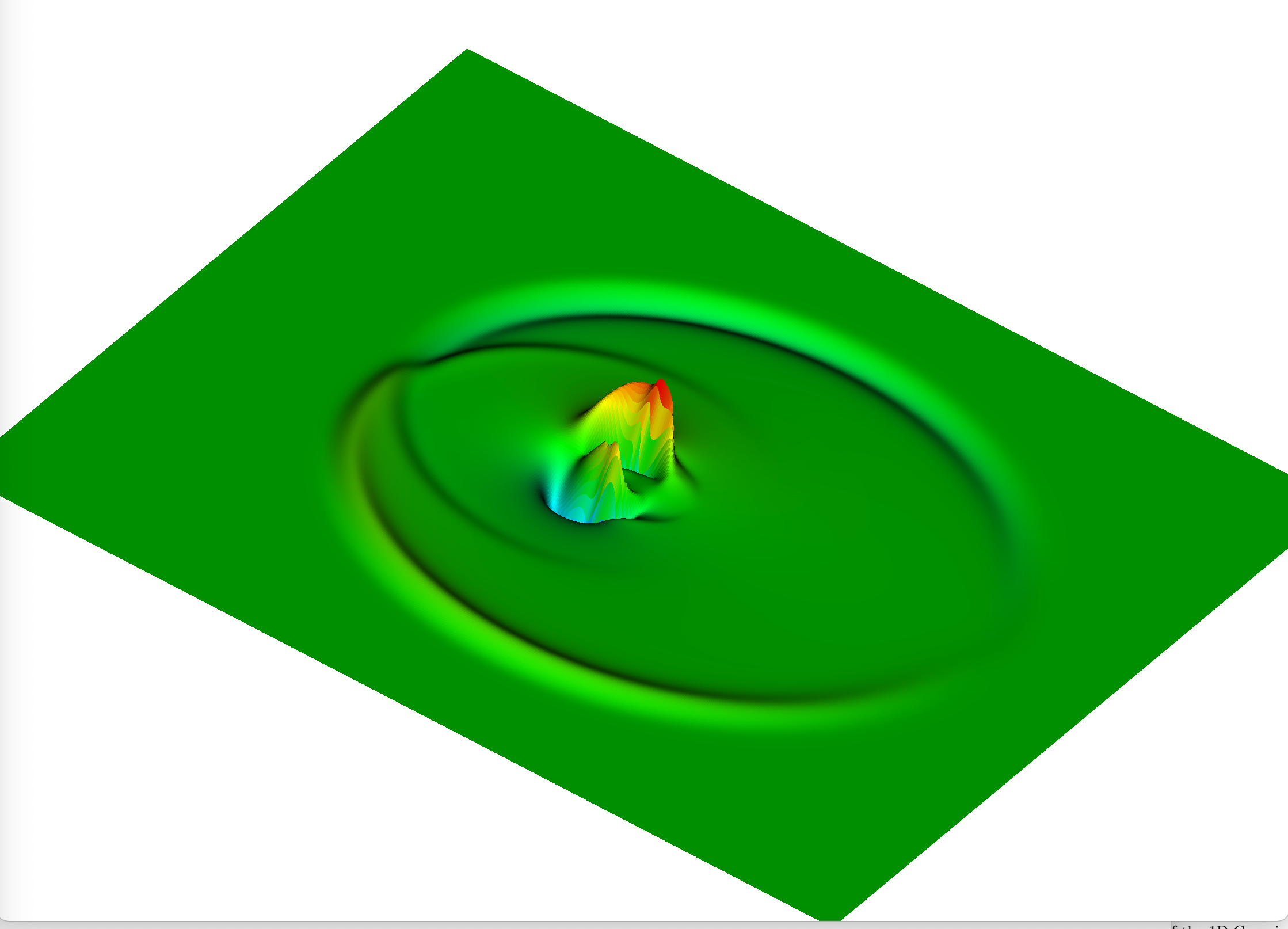}
\includegraphics[width=3.2in]{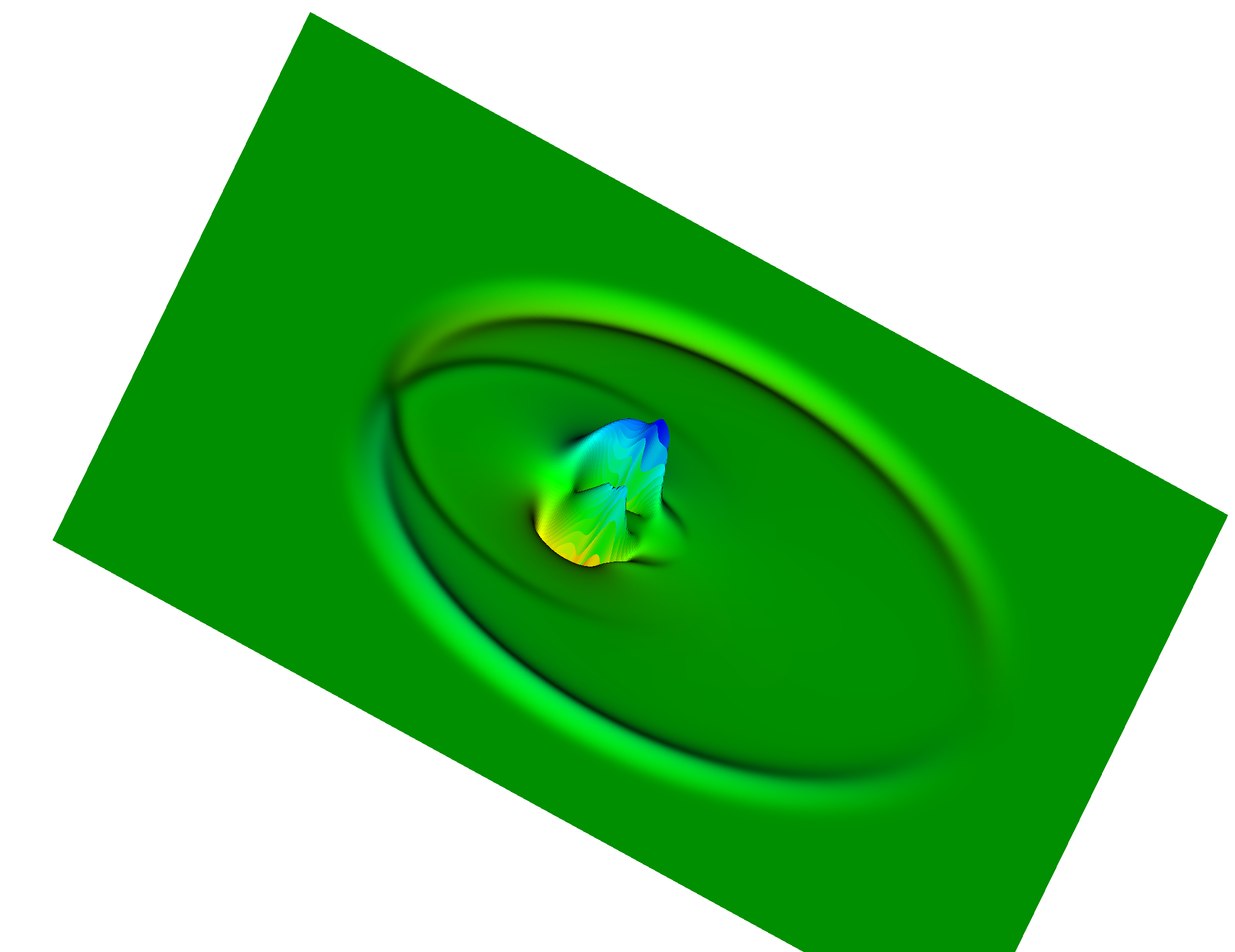}
(a)  $B_x > 0$ perspective , \quad  \quad  \quad \qquad  (b)  $B_x < 0$ perspective  - c.f.  Fig. 3
\caption{The $B_x(x,z)$ field generated from the interaction of the 1D pulse with the 2D dielectric.  Time snapshot and axes scales  corresponds to that of Fig. 3.  (a) The $B_x > 0$, and (b) $B_x < 0$.  
}
\end{center}
\vspace{0cm} 
\end{figure} 

\begin{figure}[!] \ 
\begin{center}
\vspace{0cm} 
\includegraphics[width=3.2in]{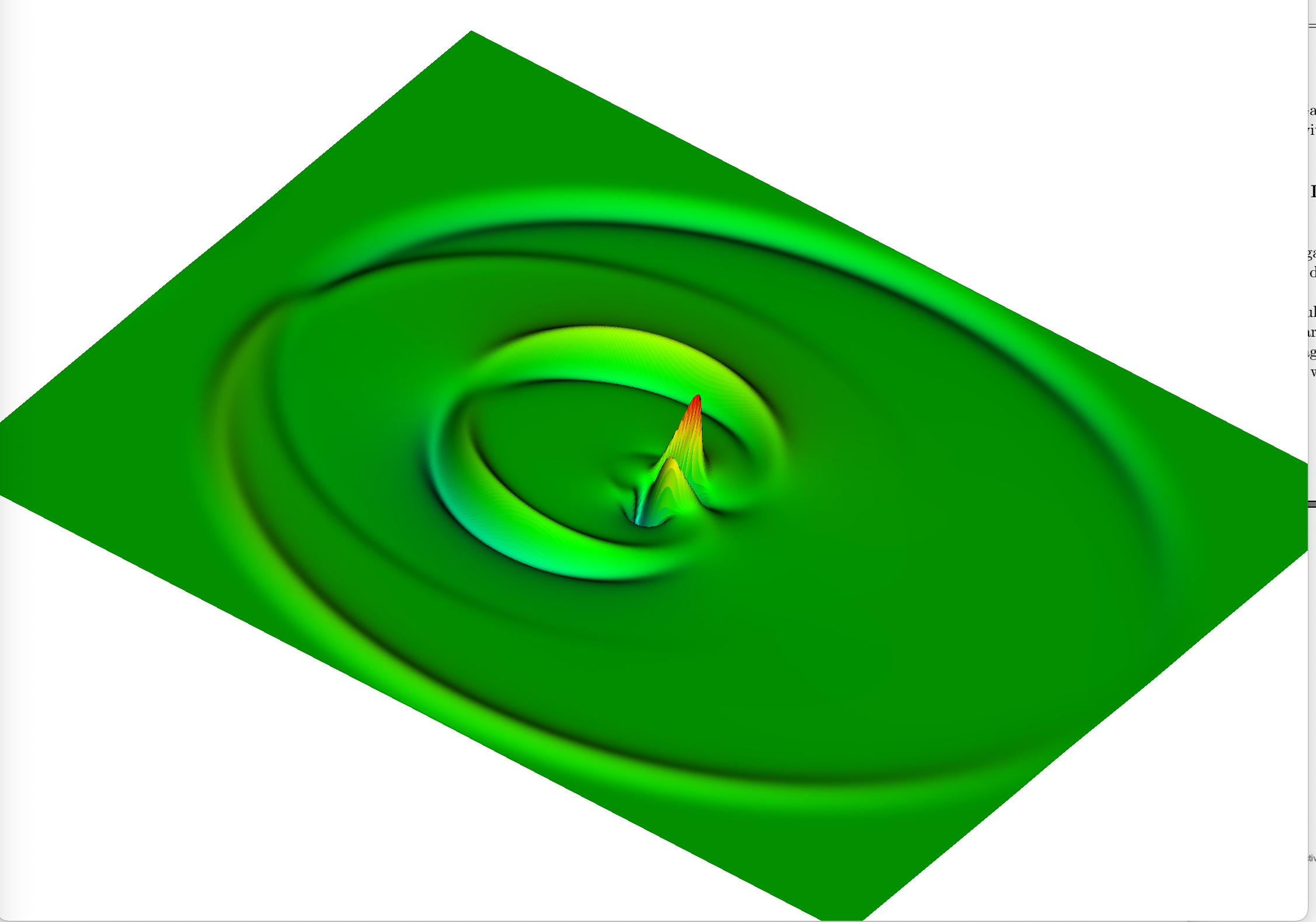}
\includegraphics[width=3.2in]{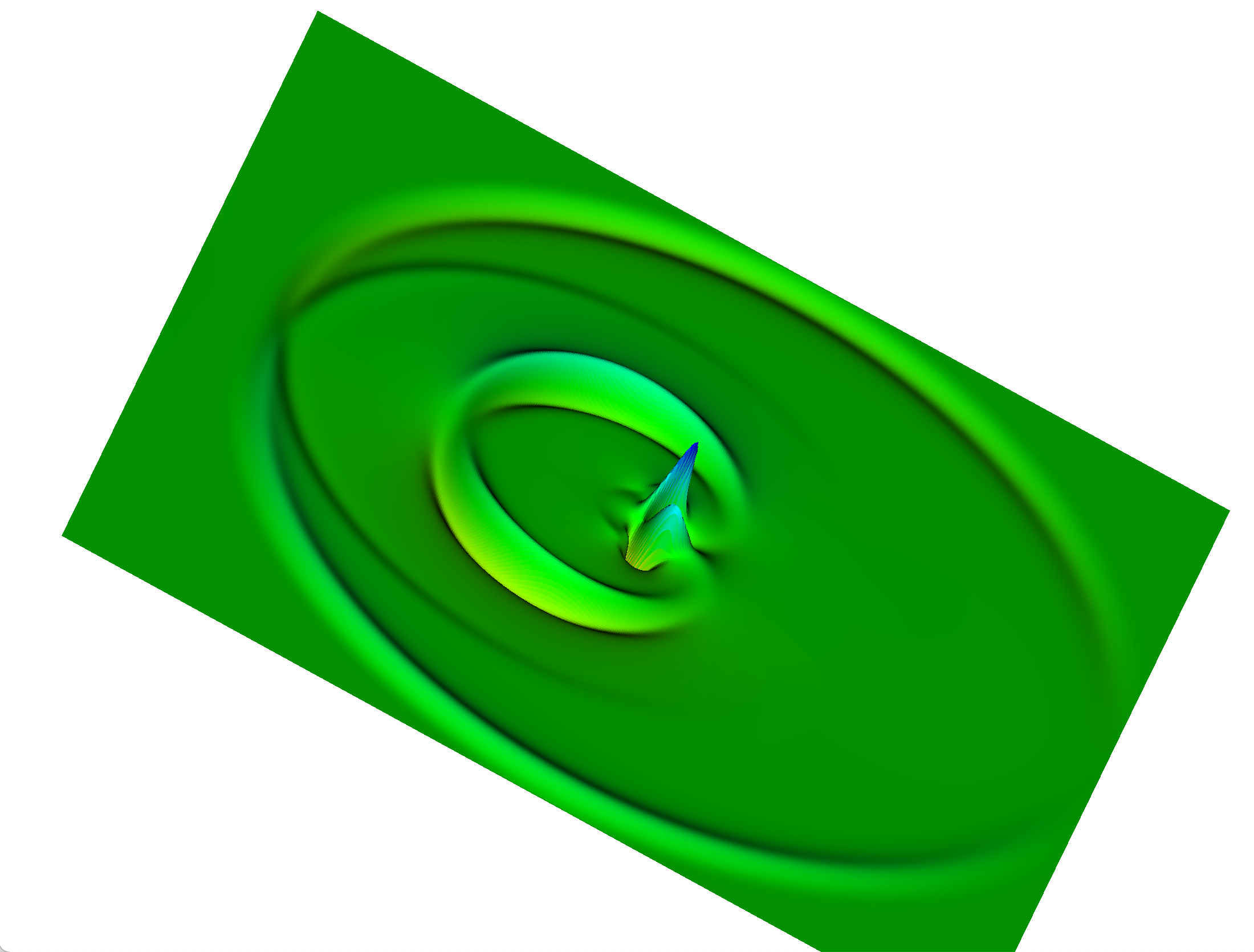}
(a)  $B_x > 0$ perspective , \quad  \quad  \quad \qquad  (b)  $B_x < 0$ perspective - c.f.  Fig. 4
\caption{The $B_x(x,z)$ field generated from the interaction of the 1D pulse with the 2D dielectric.  Time snapshot and axes scales corresponds to that of Fig. 4.  (a) The $B_x > 0$, and (b) $B_x < 0$. 
}
\end{center}
\vspace{0cm} 
\end{figure} 
The mirror symmetry of Figs. 2-7 is a $\pi$-rotation about an axis parallel to the x-axis and passing through the midpoint $z=L/2$ of the z-axis.  {The simulation box is a square of side $L$.)  From the QLA simulations we see dipole structure in the generated $B_x$-field.

In Fig. 8 we plot the
central differenced post-processed $\nabla \cdot \mathbf{B}(x,z)$ for the time snapshot corresponding to Fig. 4 and Fig. 7.
 The vertical axis scale is multiplied by a factor of $10^{-3}$ when compared to the vertical axes scales in Fig. 4 and Fig. 7.
Essentially, $\nabla \cdot \mathbf{B}$ is zero everywhere, except at the localized $B_x$ structures (see Fig. 7) where 
 $\nabla \cdot \mathbf{B}$ is done by 3 orders of magnitude over the local values of $B_x$.
\begin{figure}[h] \ 
\begin{center}
\vspace{0cm} 
\includegraphics[width=3.2in]{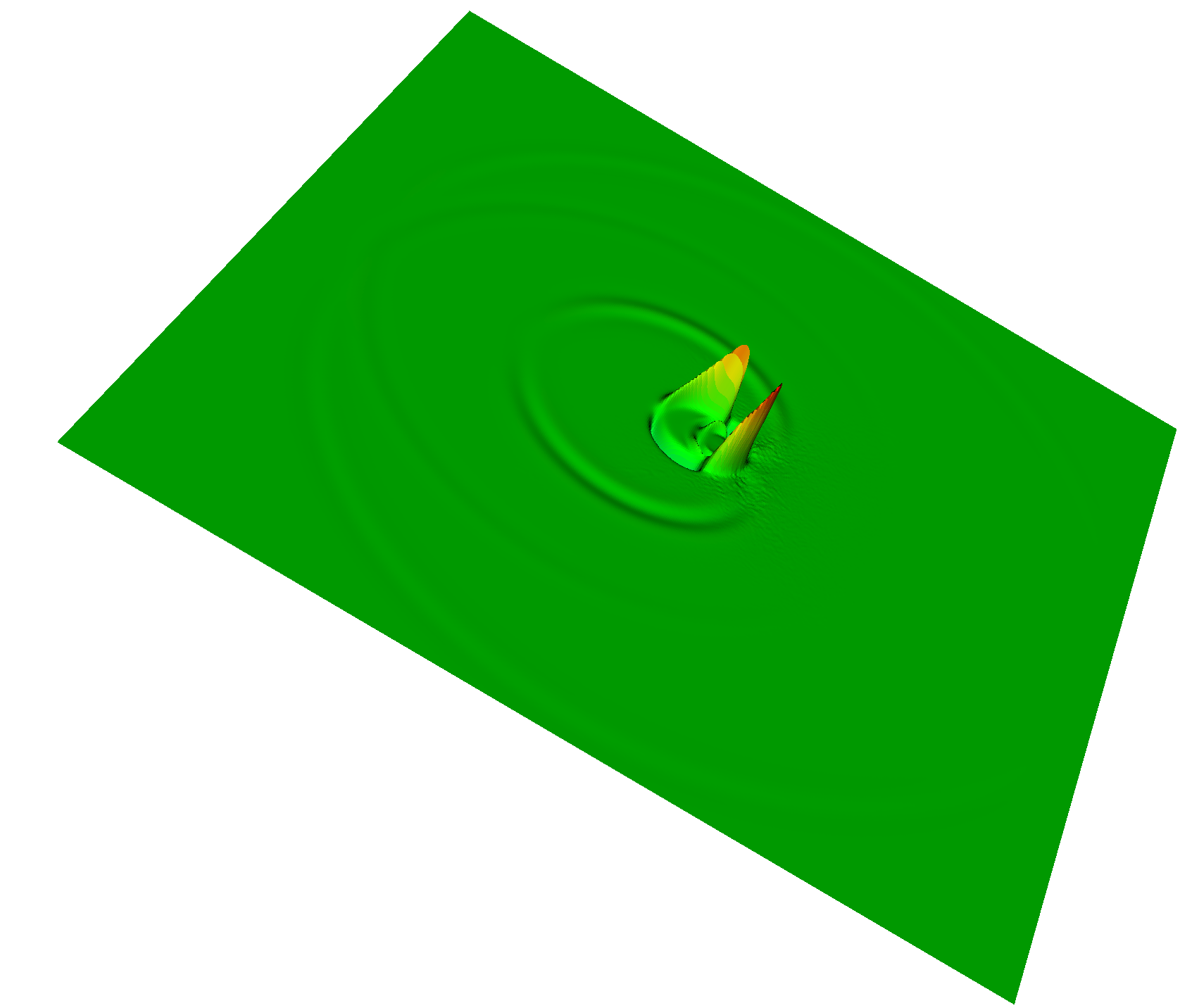}
\includegraphics[width=3.2in]{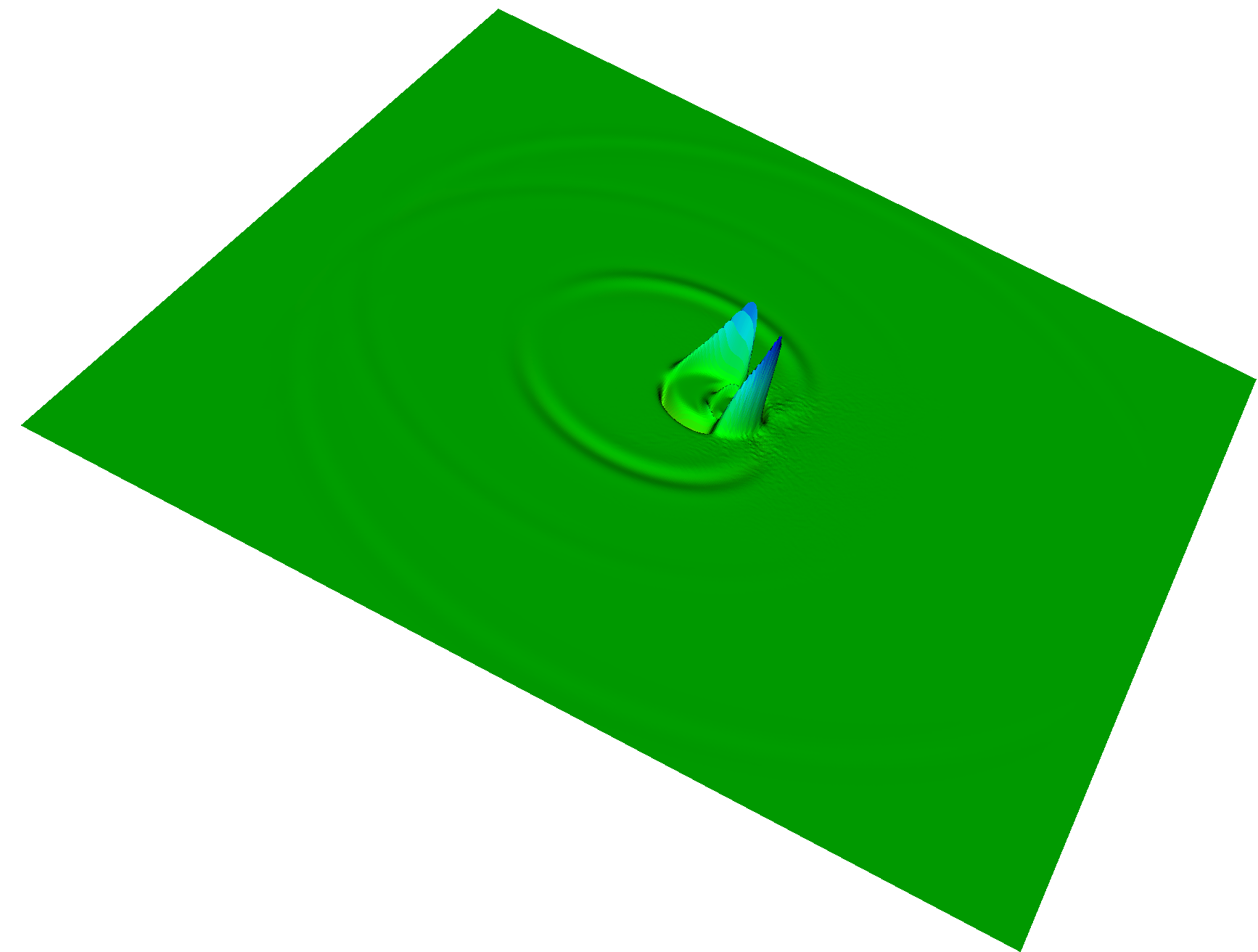}
(a)  $\nabla \cdot \mathbf{B} (x,z,t) > 0$ perspective , \quad  \quad  \quad \qquad  (b)  $\nabla \cdot \mathbf{B} (x,z,t) < 0$ perspective.
\caption{The  $\nabla \cdot \mathbf{B} (x,z,t) $ from the QLA code. Time snapshot corresponds to that of Fig. 4 and Fig. 7 - except the vertical axis should here be scaled by a factor of $10^{-3}$:  (a) $\nabla \cdot \mathbf{B} > 0$, and (b) $\nabla \cdot \mathbf{B} < 0$ 
}
\end{center}
\vspace{0cm} 
\end{figure}  
 
 \subsection{Gaussian Pulse interacting with a 2D Cylindrical Hole  :  $n_1 > n_2$ }
 In this next set of QLA simulations, we consider the 1D pulse now propagating from high refractive index $n_1 = 3$ medium onto a cylindrical vacuum hole with $n_2 = 1$.  When the pulse starts to interact with the cylindrical hole, the fields within the hole have a greater phase velocity than the pulse within the incident higher refractive index medium so that that wavefront will lead the pulse, as seen in Fig. 9(a) and (b).                                                                                                                                                                                                                                 \begin{figure}[!] \ 
\begin{center}
\vspace{0cm} 
\includegraphics[width=3.2in]{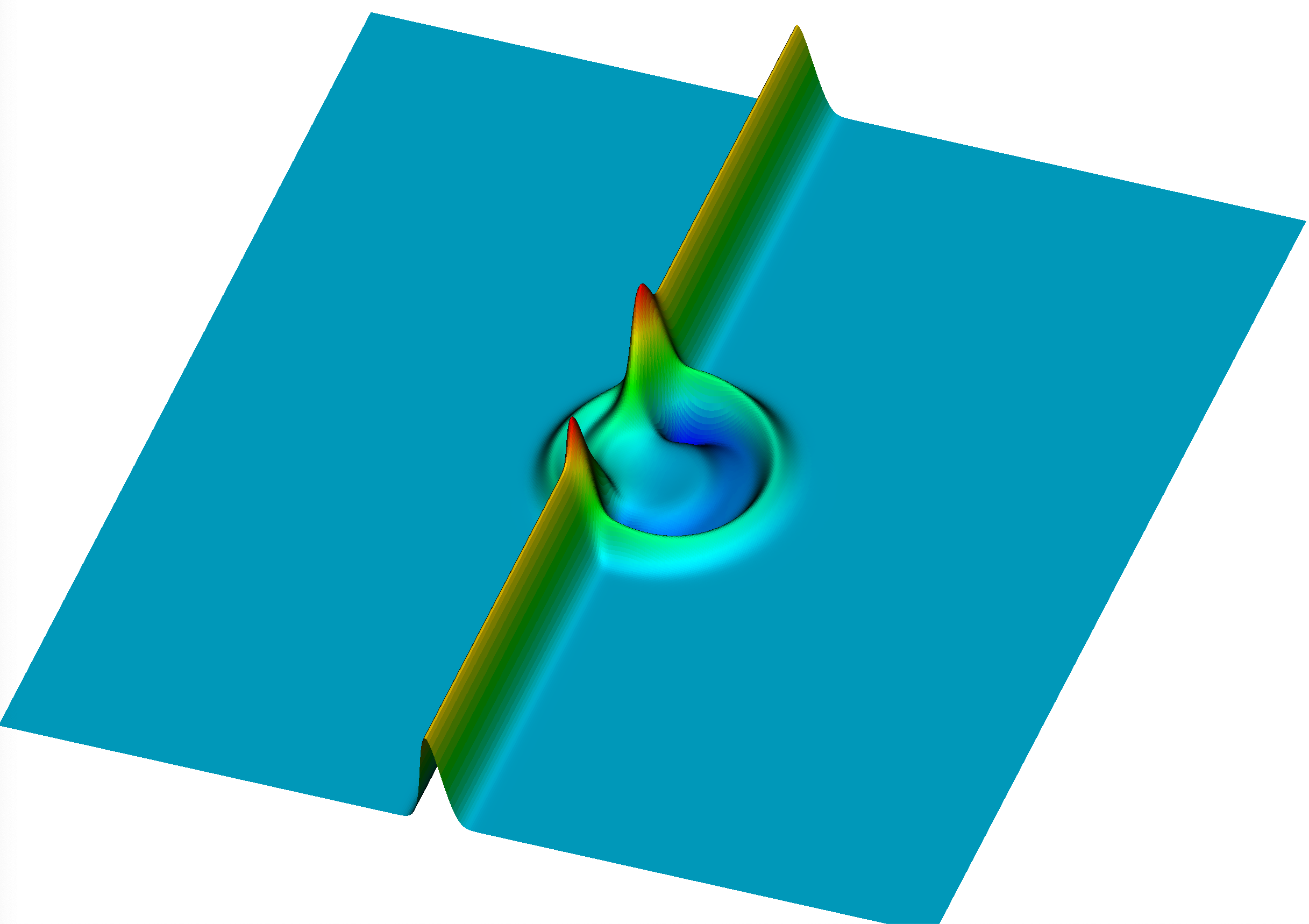}
\includegraphics[width=3.2in]{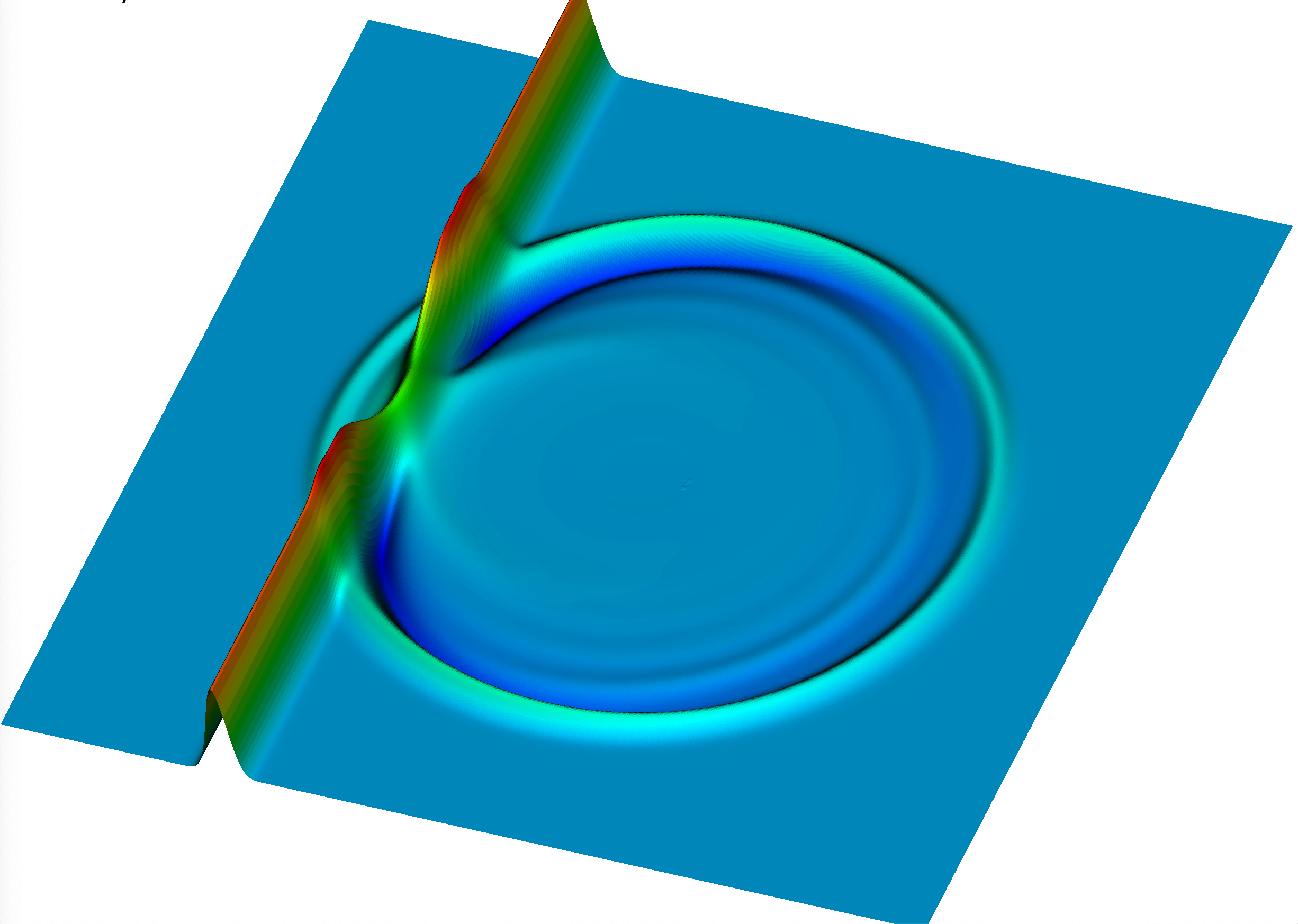}
(a)  $E_y$ field in early stages, \quad  \quad  \quad \qquad  (b)  $E_y $ field at late stages.
\caption{The electric field component $E_y$ for scattering of a pulse from $n_1$ to $n_2$, but with $n_1 > n_2$.  (a)  early stages of the scattered field from the vacuum hole, (b)  late stages of the scattered field.
}
\end{center}
\vspace{0cm} 
\end{figure}   
 Moreover, the first interaction of the pulse with the vacuum hole leads to a positive ring of reflected  electric field in contrast with the scattering from region $n_1$ into region $n_2$ with $n_1 < n_2$ where the reflected electric field ring is negative.  These signs on the reflected electric field are, of course, reminiscent to the Fresnel jump conditions in 1D.  
 
 \subsection{Gaussian Pulse interacting with a 2D Cylindrical Cone  :  $n_1 <  n_2$ } 
 We now consider QLA simulations of a Gaussian 1D pulse onto a dielectric cone, Fig 10a.
 \begin{figure}[h] \ 
\begin{center}
\vspace{0cm} 
\includegraphics[width=3.2in]{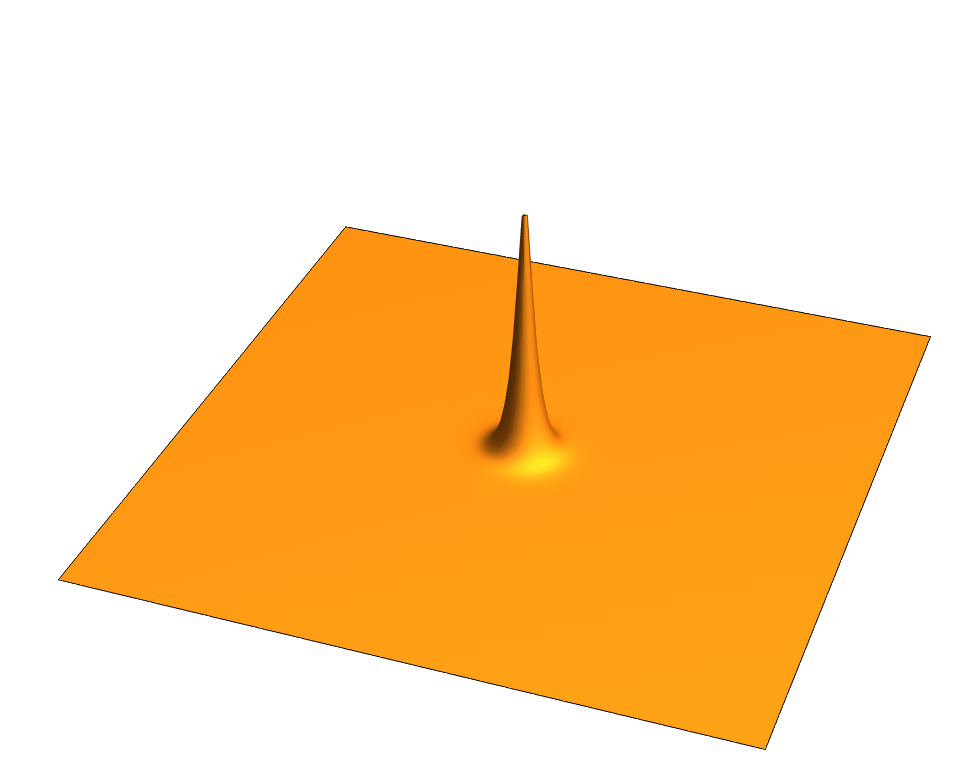}
\includegraphics[width=3.2in]{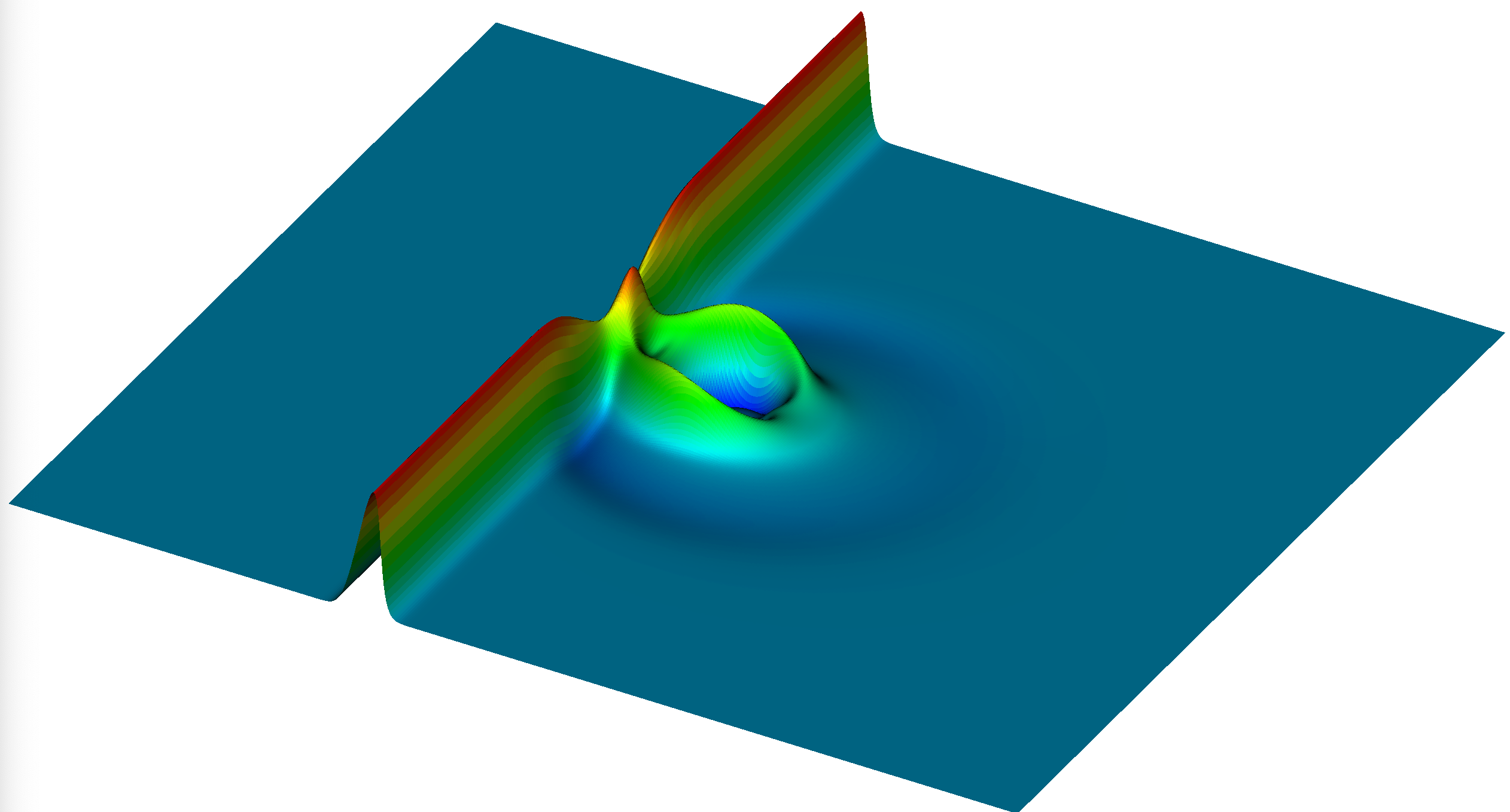}
(a)  Dielectric cone, \quad  \quad  \quad \qquad  (b)  $E_y $ field in early scattering stage from the cone.
\caption{(a)  The dielectric cone, with apex having $n_2 = 3$, (b)  $E_y$ in early stages of interacting with the dielectric cone.  
}
\end{center}
\vspace{0cm} 
\end{figure}   
 Because the refractive index gradients are weak at the edges of the cone, there is no prominent reflected ring of negative electric field, Fig 10(b),  as in the case of the sharp edge gradients in the dielectric cylinder case, Fig. 2(a).  Later in time, Fig. 11, one sees only a secondary ring of predominantly $E_y > 0$ , Fig. 11(a) and (b).
 
 \begin{figure}[!] \ 
\begin{center}
\vspace{0cm} 
\includegraphics[width=3.2in]{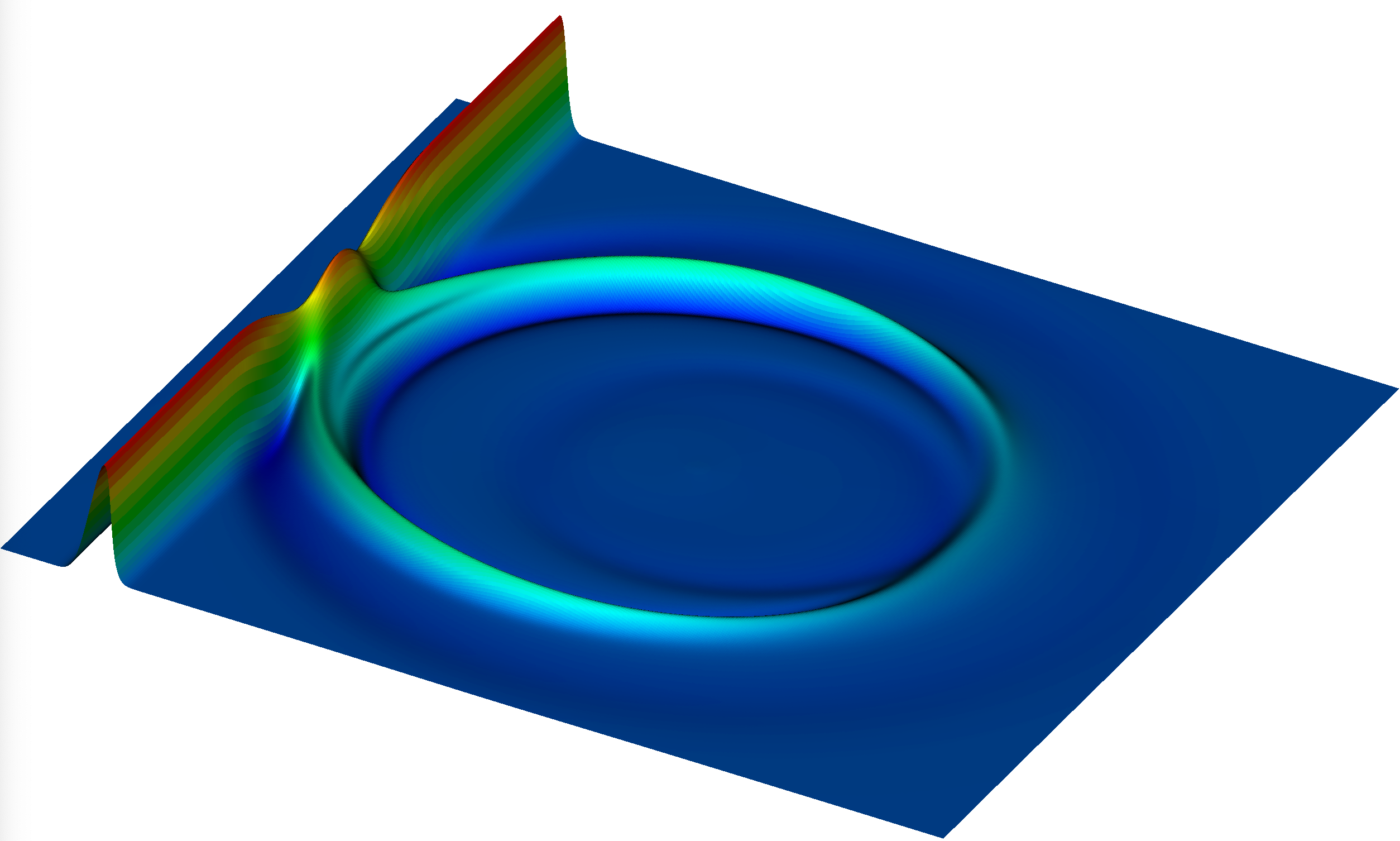}
\includegraphics[width=3.2in]{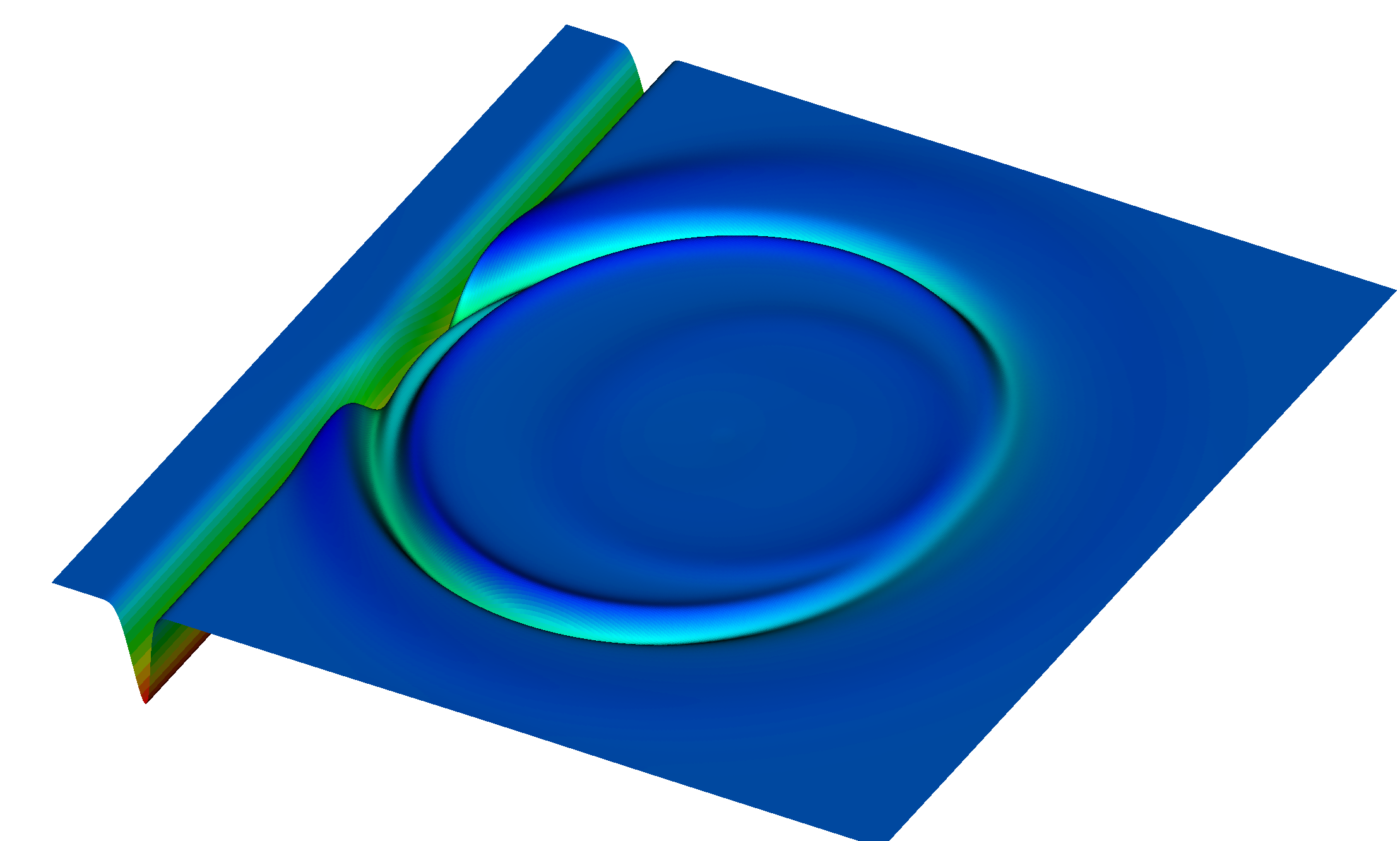}
(a)  $E_y$ field - perspective of $E_y > 0$. \quad  \quad  \quad \qquad  (b)  $E_y $ field but now from the perspective $E_y < 0$.
\caption{At late stage in the interaction of the pulse with the dielectric cone, one does not see strong internal wavefronts being emitted from the dielectric cone.  This should be compared to the emitted wavefronts from the dielectric cylinder, Fig. 4.
}
\end{center}
\vspace{0cm} 
\end{figure} 

Plots of $\nabla \cdot \mathbf{B}(x,z,t)$ for the scattering of the pulse from dielectric cones are similar to that for scattering from the cylindrical dielectric, Fig. 8, except the peak magnitudes are typically now reduced by a further factor of $50$ over the sharp dielectric boundary layer of the vacuum-cylinder interface .

 \subsection{The role of the collision angle perturbation parameter $\varepsilon$ } 

in order to develop an explicit QLA we must introduce a perturbation parameter $\varepsilon$ into the collision and potential operators.  Here, we have developed a 2D QLA for Maxwell equations in a scalar dielectric medium that recovers the Maxwell equations to second order in $\varepsilon$.  From the simulations we can identify $\varepsilon$ to be the scaled speed of propagation of the pulse in a vacuum.   In our previous QLA for the nonlinear Schrodinger equation and for the (3D) spinor Bose-Einstein condensate equations, $\varepsilon$ was related to the soliton/quantum vortex amplitude. For these nonlinear equations, the simulations of the QLA models revealed an upper bound to $\varepsilon_{max}$ above which the QLA no longer recovered the appropriate soliton/quantum vortex time behavior. This is the well known price we are forced to pay when applying perturbation theory.  

However, for our QLA for Maxwell equations we find that the QLA simulations faithfully reproduce the Maxwell equations even for $\varepsilon = 1.0$, the maximal possible speed of pulse propagation.  In Fig 12 - 14 we show the time evolution snapshots of the $E_y$-field for two different values of $\varepsilon$ :  $\varepsilon= 0.01$ and $\varepsilon= 1.0$.  Of course, the computational time scales as $1/ \varepsilon$.

\begin{figure}[h] \ 
\begin{center}
\vspace{0cm} 
\includegraphics[width=3.2in]{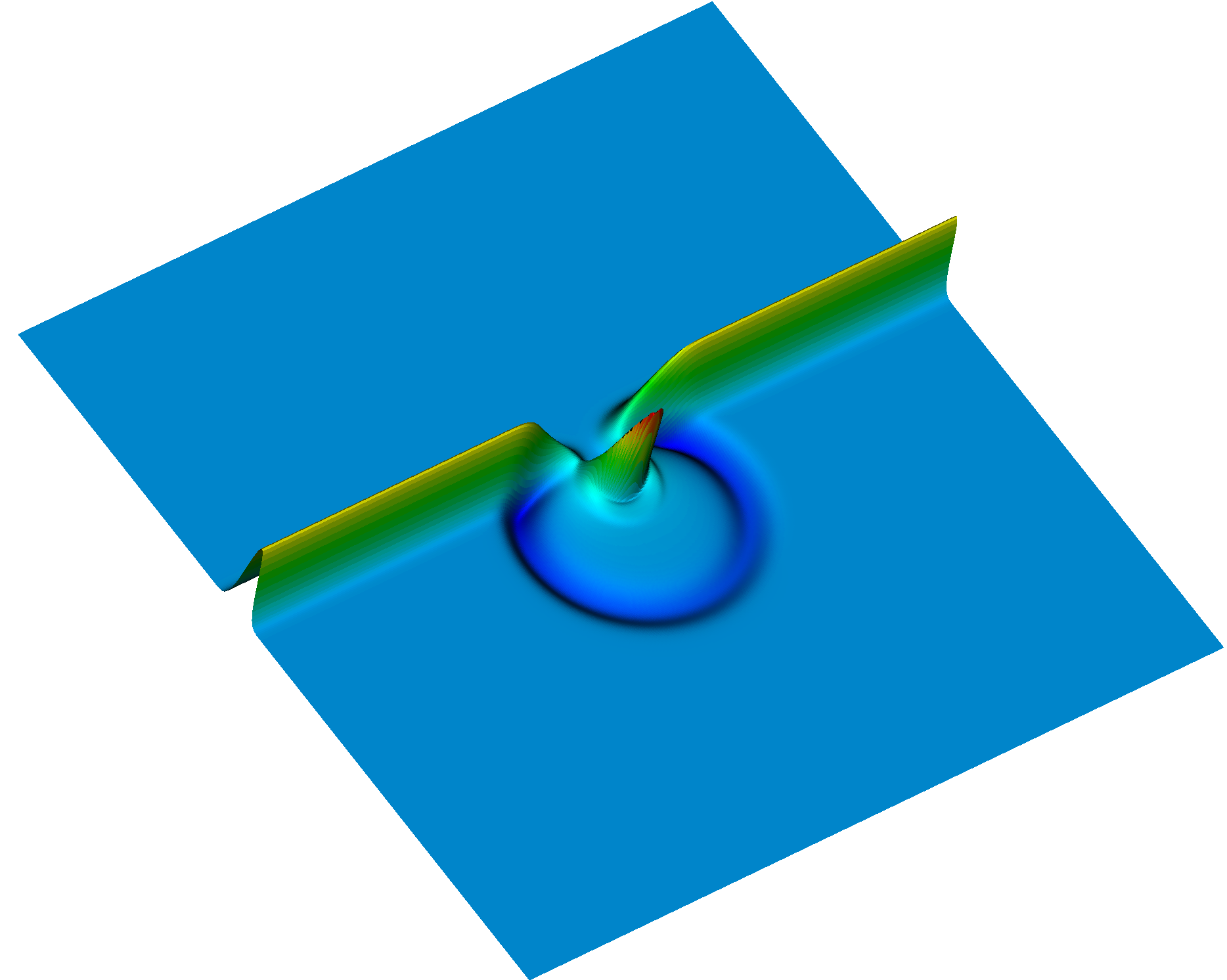}
\includegraphics[width=3.2in]{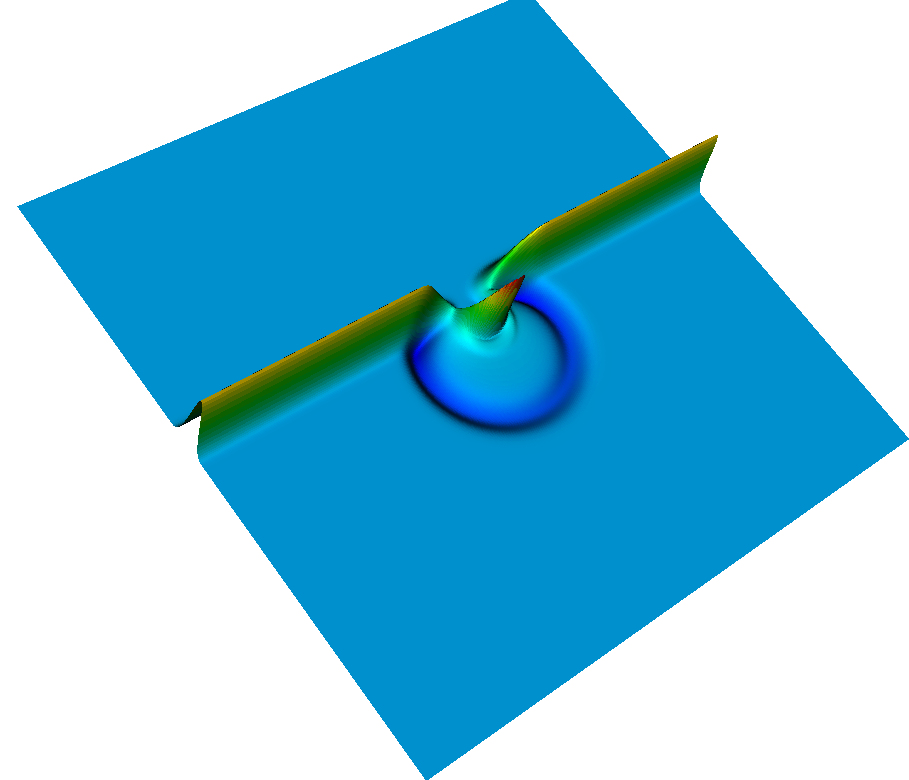}
(a)  $E_y$ field for $\varepsilon = 0.01$. \quad  \quad  \quad \qquad  (b)  $E_y$ field for $\varepsilon = 1.00$
\caption{The effect of the QLA perturbation parameter $\varepsilon$ on the evolution of the $E_y$ field soon after the pulse has interacted with the dielectric cylinder.
}
\end{center}
\vspace{0cm} 
\end{figure} 

 \begin{figure}[!] \ 
\begin{center}
\vspace{0cm} 
\includegraphics[width=3.2in]{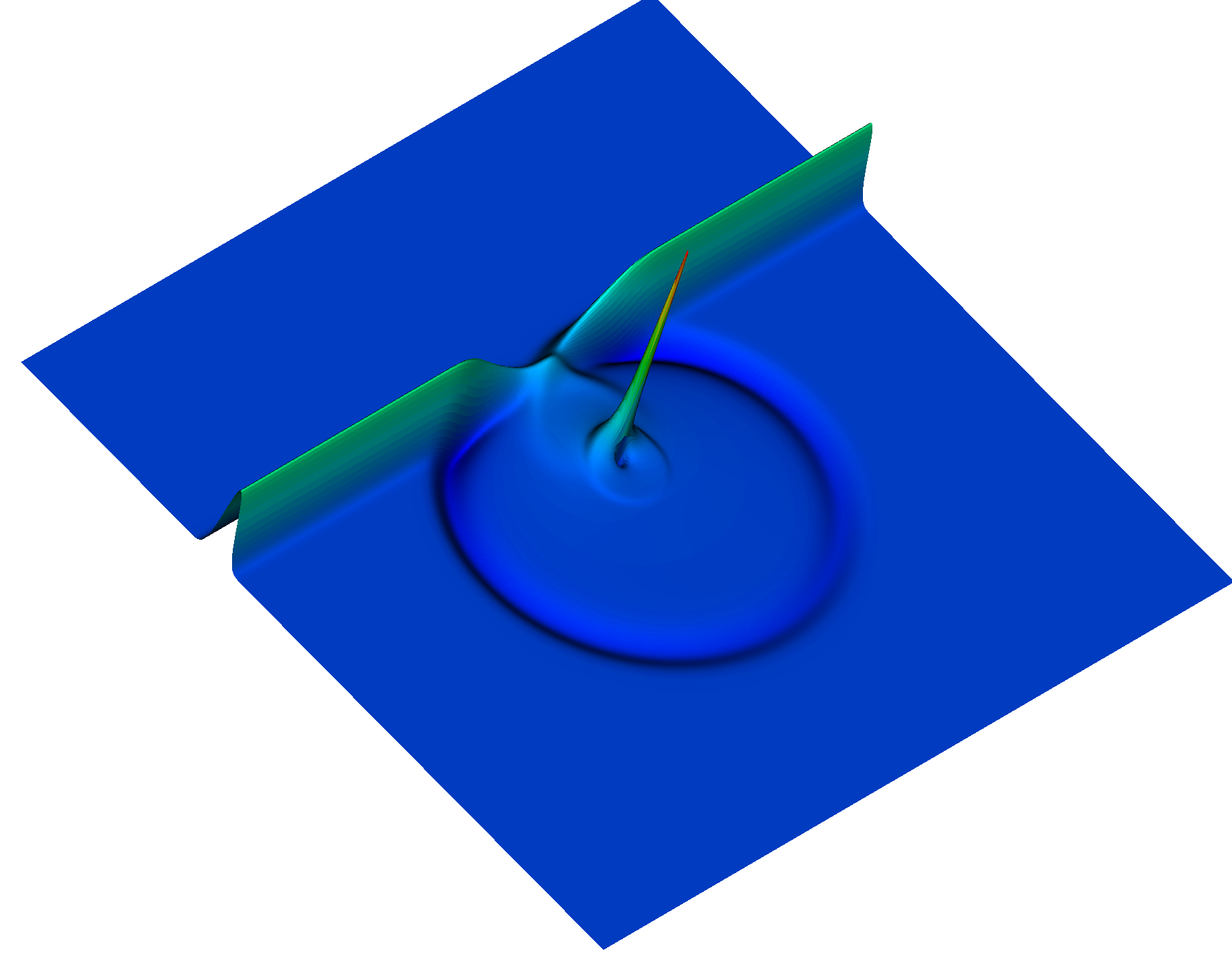}
\includegraphics[width=3.2in]{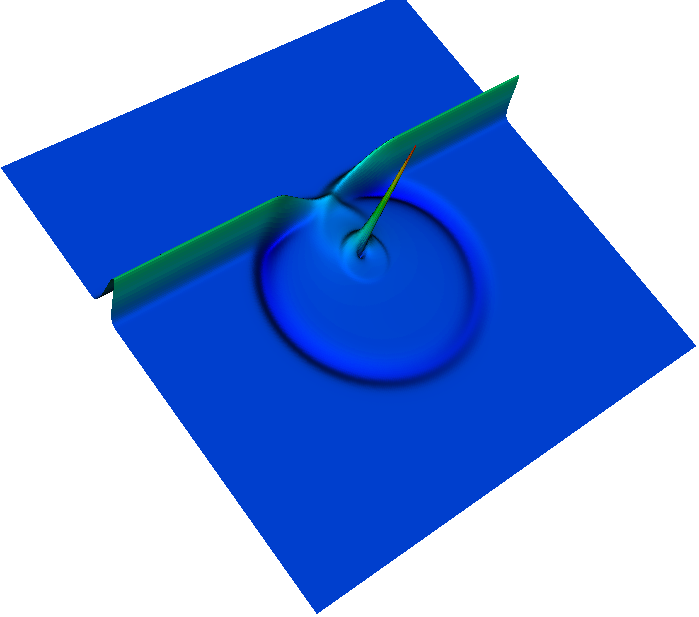}
(a)  $E_y$ field for $\varepsilon = 0.01$. \quad  \quad  \quad \qquad  (b)  $E_y$ field for $\varepsilon = 1.00$
\caption{The development of $E_y$ with its peak  (a) $0.0330$ , while for (b) $0.0323$
}
\end{center}
\vspace{0cm} 
\end{figure} 

 \begin{figure}[!] \ 
\begin{center}
\vspace{0cm} 
\includegraphics[width=3.2in]{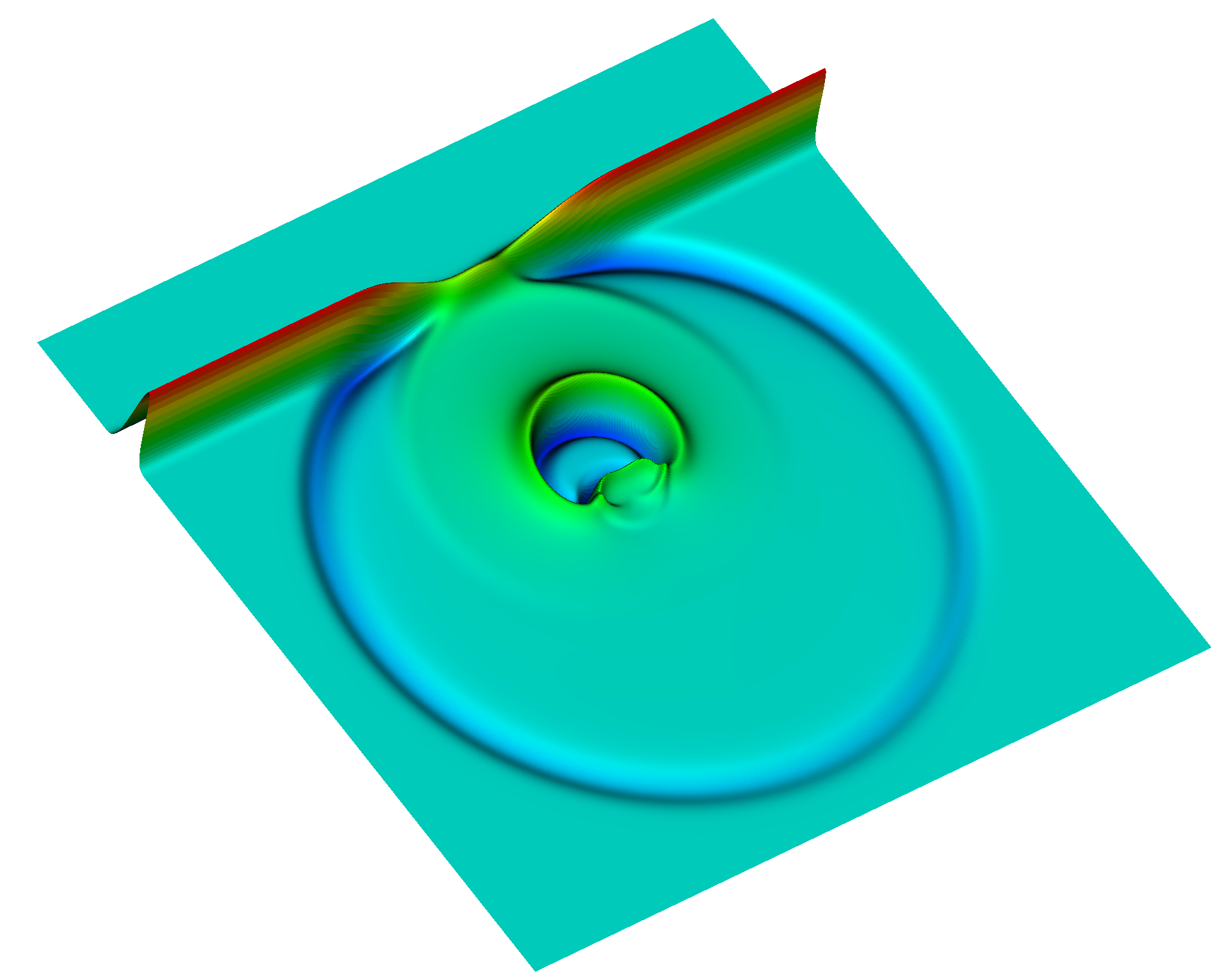}
\includegraphics[width=3.2in]{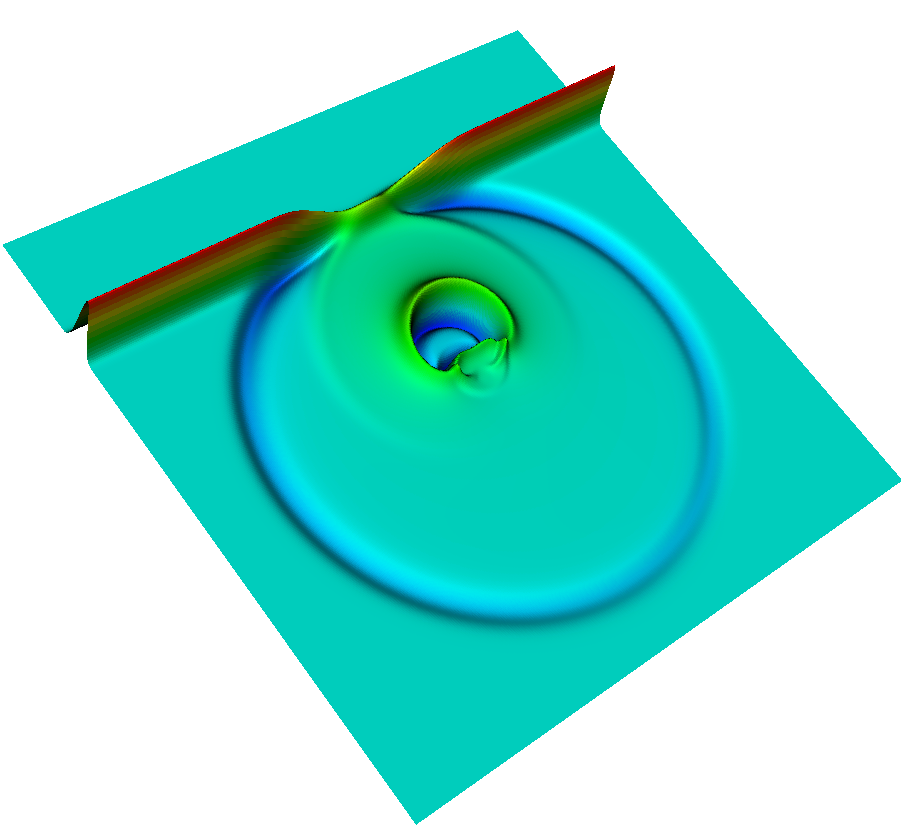}
(a)  $E_y$ field for $\varepsilon = 0.01$. \quad  \quad  \quad \qquad  (b)  $E_y$ field for $\varepsilon = 1.00$
\caption{The late stage in the evolution of $E_y$ following its interaction with the dielectric cylinder
}
\end{center}
\vspace{0cm} 
\end{figure} 

\section{Summary and Conclusions}

\qquad  We have performed detailed QLA simulations of the scattering of an electromagnetic pulse off a localized 2D scalar dielectric medium.  If the strength of the spatial gradient of the dielectric boundary layer is large, we find significant reflections/transmissions of the pulse within the dielectric.  This leads to the dielectric emitting outgoing wavefronts into the vacuum like a quasi-antenna.  

In our current formulation of the Maxwell equations, we are only solving $\nabla \cdot \bf{B} = 0$ to second order in the unitary collision angle parameter $\varepsilon$.  Nevertheless we do not seem to be tainted by this in the simulations.  A possible reason for this is that QLA is a mesoscopic representation and the Maxwell equations are recovered in the continuum limit.  Hence the time evolution of the QLA simulation does not deal directly with the magnetic field $\bf{B}$.  

QLA's are necessarily perturbative theories since only in extremely rare instances can one perform a closed form representatioin of the collision matrix.  In our earlier work on solving the 1D Nonlinear Schrodinger equation $(NLS)$ or the 2D and 3D Gross-Pitaevskii equation for the evolution of a Bose-Einstein condensate $(BEC)$ ground state or the further generalizations to spinor BECs, the perturbation parameter in the collision operators was related to the amplitude of the wave function.  The QLA recovers the desired continuum equations perturbatively, and QLA simulations have determined an $\varepsilon_{max}$ above which the QLA simulations do not yield solutions to the desired continuum equations.  For example, in 1D NLS, one recovers the exact soliton-soliton collision features for $\varepsilon < \varepsilon_{max}$.  However, if we choose $\varepsilon > \varepsilon_{max}$ one sees the soliton breakdown in time.  All this is saying is that the QLA is now representing an equation different from 1D NLS - because of the choice of too high an $\varepsilon$.  In our QLA representation of the Maxwell equations in scalar media, we find that the $\varepsilon$ is just the normalized speed of light in that medium.  Our simulations show that the QLA holds for Maxwell equations even for $\varepsilon = 1$.

Finally QLA can be viewed from 2 sides:  on the one side it is basically a unitary qubit lattice representation of Maxwell equations which should be able to be run on error-correcting quantum machines as they come on-line.  It has been noticed by Khan [28] that under the Riemann-Silberstein-Weber representation, Eq. (3), the structure of Maxwell equations in a scalar dielectric medium are non-unitary:  the evolution operators that involve the derivative on the dielectric function are Hermitian, and not unitary.  In our QLA  this is seen in our potentital collision operators being non-unitary.  This will make the application of qubit algorithms for plasma physics to be more challenging since clever constructs need to be performed in order to still represent such non-unitary systems on a quantum computer [30] or else seek different representations of the fields $\mathbf{E}$ and $\mathbf{B}$.  Similar comments can be made on how to represent nonlinear physics on a linear quantum mechanical computer.  
The other way of looking at QLA is that it is an ideally parallelized algorithm for running on classical supercomputers.  Indeed our QLA spinor BEC codes did not see saturation with nodes on the IBM $Mira$ machine, even up to the maximum $786,432$ cores available to us (in 2015).  This is also why we have run on such large grids for our Maxwell equations:  it is basically to test out and tune the algorithm as we move to harder and harder problems.

\section{Acknowledgments}
This research was partially supported by Department of Energy grants DE-SC0021647, DE-FG02-91ER-54109, DE-SC0021651, DE-SC0021857, and DE-SC0021653.  The simulations were performed on $CORI$ of the National Energy Research Scientific Computing Center (NERSC), A U.S. Department of Energy Office of Science User Facility located at Lawrence Berkeley National Laboratory, operated under Contract No. DE-AC02-05CH11231.

\section{\textcolor{black}{APPENDIX: Additional Comments on Qubit Lattice Algorithms}}
\subsection{\textcolor{black}{QLA and Fresnel Conditions at Interfaces}}
{\textcolor{black}{An unexpected result from our 1D QLA simulations is in the 1D transmission/reflection of an electromagnetic pulse from a dielectric interface.  
The textbook Fresnel conditions on the electric field amplitudes are derived for a $\delta$-function interface 
assuming  an incident plane wave, a reflected plane wave and a transmitted plane wave co-existing at the interface.
The boundary conditions at this interface yield $E_{refl}/E_{inc} = (n_1 - n_2)/(n_1 + n_2)$
while $E_{trans}/E_{inc} = 2 n_1/(n_1+n_2)$.}

\textcolor{black}{In QLA simulations the  initial value problem follows the evolution of a pulse onto a thin continuous refractive index  layer that smoothly joins the $n_1$ to the $n_2$-dielectrics.
Thus the QLA solves the Maxwell equations throughout all space, without the imposition of any internal boundary layers.  This is quite distinct from 
standard computational methods like Finite-Difference Time-Domain (FDTD), which do impose internal boundary conditions at the dielectric interface.}

{\textcolor{black}{While we [29] have developed a theory to explain this QLA result,  we will show by a simple choice of initial pulse this factor of $\sqrt{n_2/n_1}$ is necessary to satisfy conservation of energy.  Consider an electromagnetic rectangular pulse propagating in the x-direction, Fig. 15a.   The steepness of the rectangular spatial end points is chosen to keep the Gibbs oscillations at an acceptable level.  The rectangular pulse propagates from a vacuum region, $n_1 = 1$ into a dielectric region with refractive index $n_2 = 2$, located at $4000 < x < 6000$.  
For an initial pulse centered at $x=2000$, with pulse width $\Delta_0 = 2000$,  the Poynting flux 
\begin{equation}
S(t)= \int_0^L \mathbf{E}(t) \times \mathbf{ B}(t) \cdot \mathbf{\hat{n}} \;dx
\end{equation}
is $S(0) = 0.19603$.  A simple estimate yields $S^{\prime}(0) \approx n_1 \Delta_0 E_{inc}^2 = 0.2$, with $E_{inc} = 0.01$.}

{\textcolor{black}{In the quasi-asymptotic QLA state at $t=14000$, Fig. 16a, the reflected QLA Poynting flux $S_{refl} = 0.021839$, while a quick estimate yields $S_{refl}^{\prime} \approx n_1 \Delta_0 E_{refl}^2$.  From the QLA simulation,  $E_{refl} = -0.00334$, and estimating the reflected width $\Delta_1 = 2000$, we obtain the rectangular area $S_{refl}^{\prime}  \approx 0.0223$.  The reflected plane wave boundary value Fresnel results agree completely with the reflected QLA simulations:  $E_{refl} = -1/3$ with the reflected pulse width equal to its incident width.} 

{\textcolor{black}{The transmission Fresnel boundary value conditions yield:  in the $n_2=2$ region, $E_{trans} = 2 n_1/(n_1+n_2) = 2/3$ and the pulse width $\Delta_2 = \Delta_0  n_1/n_2  = 1000$, so that the Fresnel transmitted Poynting flux $S_{trans}^{Fresnel} = 0.0889$.  Thus the Fresnel total Poynting flux at $t=14000$ is $S_{Fresnel} = 0.0223 + 0.0889 = 0.1112$, which is very different from the initial Poynting flux $S(0) = 0.19603$.}

{\textcolor{black}{On the other hand, the transmitted quasi-asymptotic QLA  Poynting flux from simulations at $t=14000$ yields $S_{trans}^{QLA} = 0.17411$.  Adding this to the reflected QLA Poynting flux $0.021839$ yields a total QLA flux of $0.19595$ - in excellent agreement with the initial flux $S(0) = 0.19603$.  
Further, if we estimate this transmission flux integral by its rectangular area: $S_{trans}^{\prime} = n_2 \Delta_2 E_{trans}^2$, with QLA $E_{trans} = 0.009408$, we obtain $S_{trans}^{\prime} \approx 0.1770$.  This together with the reflected Poynting flux yields excellent agreement with the initial Poynting flux and thus conservation of energy. }

{\textcolor{black}{Finally, we comment on the Poynting flux during the transient overlapping times when the electromagnetic field profiles are distorted and not simple rectangular
pulses (see Fig. 15b).  As discussed in Ref 29, we must be careful in determining the reflected Poynting flux, since for $x < 4000$ we have to determine which part of the fields are incident (with unit normal $\mathbf{\hat{n}}$ in the $+ \mathbf{\hat{x}}$ direction) and which part is reflected ($\mathbf{\hat{n}}$ in the $- \mathbf{\hat{x}}$ direction) in evaluating Eq. (43).  Thus at a particular time like $t = 6000$, Fig 15(b), we rerun the QLA for refractive index $n_1$ for all x, yielding the incident fields $E_{y,n_1}(x,t=6000)$ and $B_{z,n_1}(x,t=6000)$ for $0 < x < 4000$.  On subtracting these incident fields from the full QLA simulations for the two dielectrics, we obtain the reflected field components for $0 < x < 4000$.  
In Fig. 16b, we plot the Poynting flux $S(t)$, Eq. (43).  For $t < 3000$ the incident pulse is fully in medium $n_1$, and the Poynting flux is constant to $9$ - significant digits.  This indicates that the QLA is very accurately modeling Maxwell equations in a uniform dielectric. 
Similarly, for $t > 10500$, the transmitted and reflected pulses are quasi-asymptotically stable with no overlap with the dielectric boundary layer around $x = 4000$.  Again we find for $t > 10500$ that the total Poynting flux $S(t)$ is constant to 9 digits.  For $0 < t< 15000$ , $S(t)$ is conserved  with variations in the 5th digit.  In the overlap time interval, $3500 < t < 10500$ - which has no counterpart in the plane wave Fresnel boundary value problem - the Poynting flux is conserved to 4 digits.}

\begin{figure}[!] \ 
\begin{center}
\vspace{0cm} 
\includegraphics[width=3.2in]{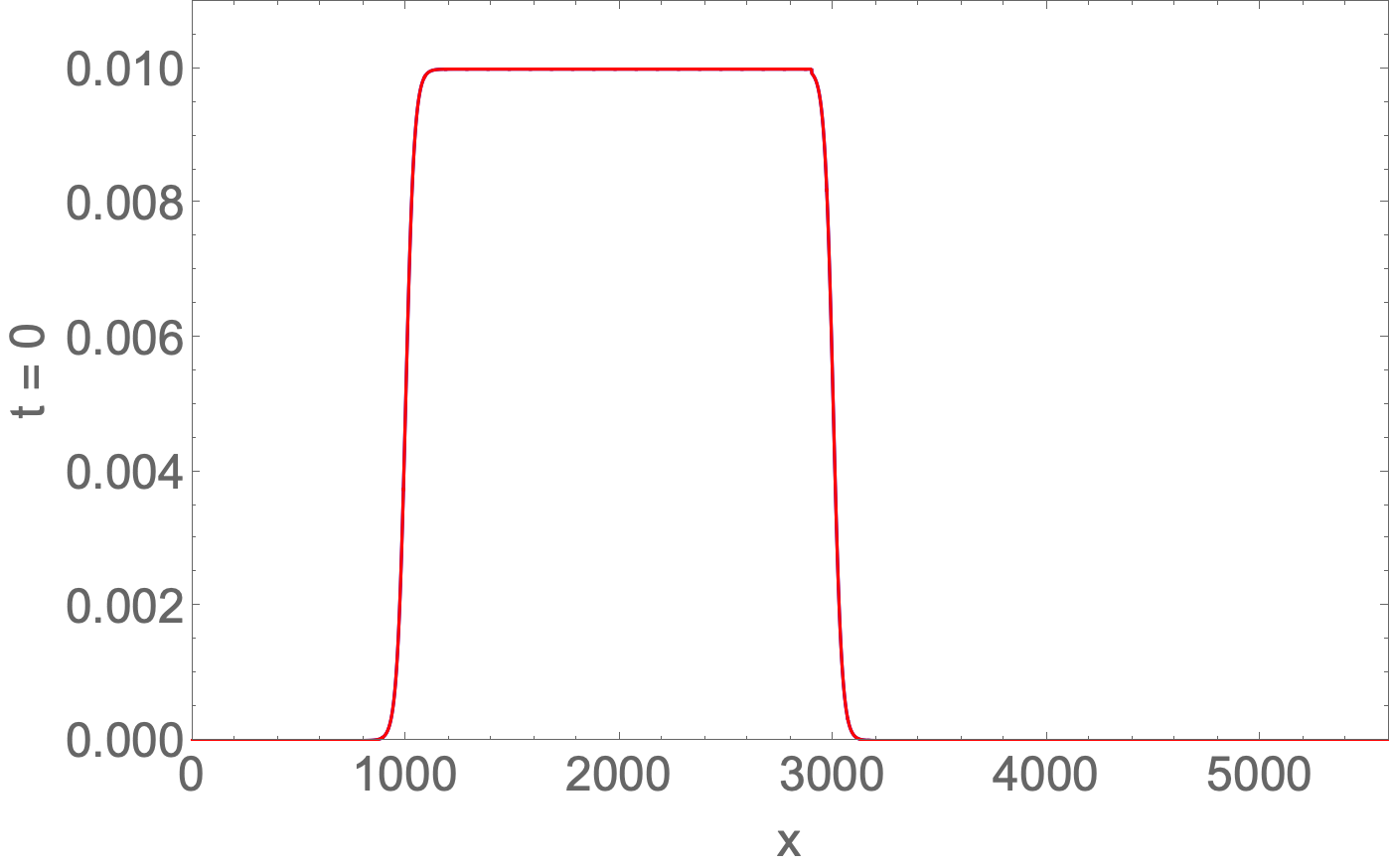}
\includegraphics[width=3.2in]{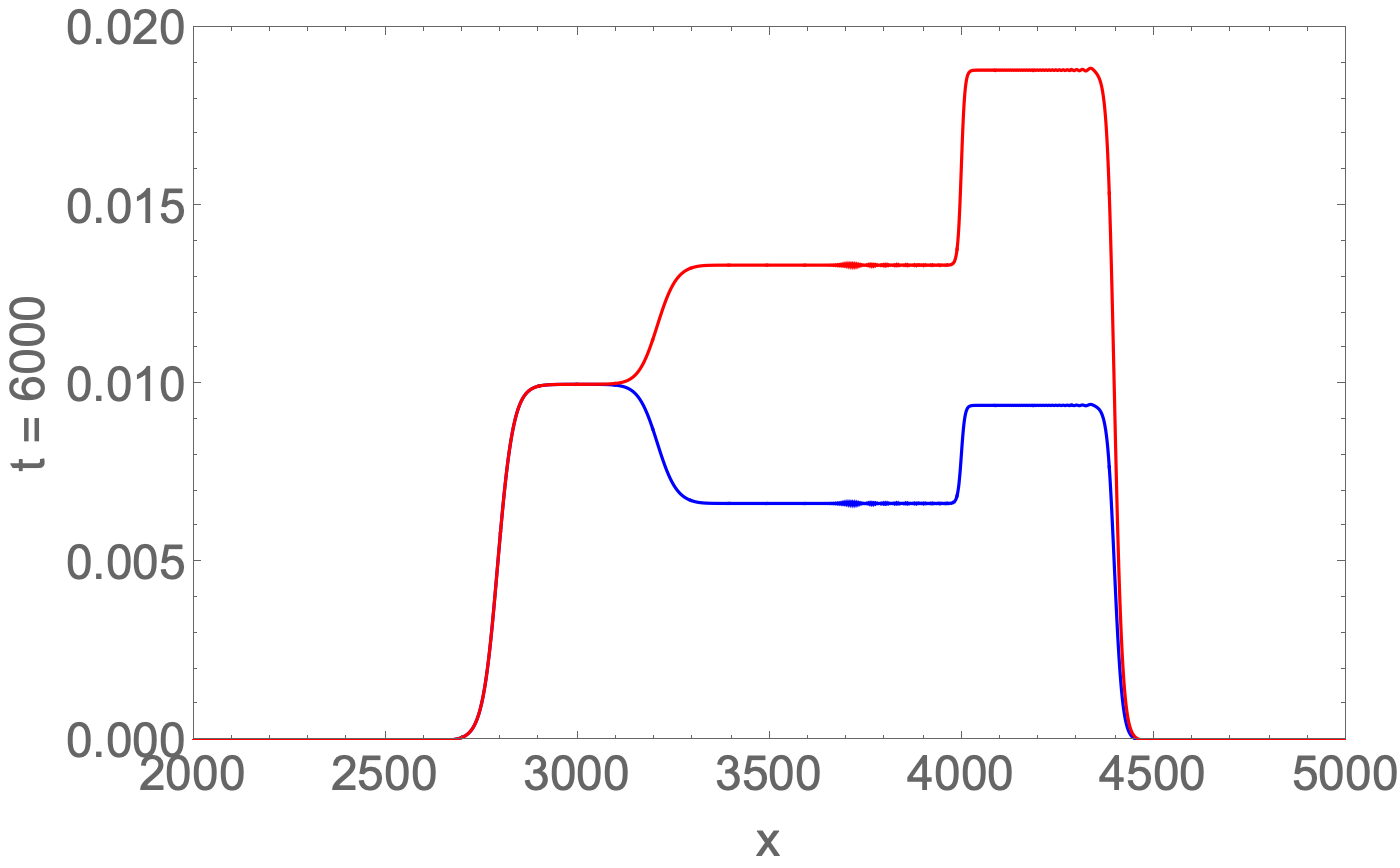}
\textcolor{red}{(a) $E_y(x,t=0 )= B_z(x,t=0)$ . \quad  \quad  \quad \qquad  (b)  $E_y $ and $B_z$ fields.}
\caption{\textcolor{black}{(a) In our QLA normalization, the initial profiles are identical $E_y(x,t=0 )= B_z(x,t=0)$ in the vacuum region, $x < 4000$, where  $n_1 = 1$.  The rectangular pulse is around 2000 lattice units in length.
(b)  At $t = 6000$ time step. there is an overlap of the rectangular pulse with the dielectric boundary layer (which is about 50 lattice units in length).  For $x < 3100$, $E_y = B_z$.  In the spatial region $3100 < x < 4000$ one has both incident and reflected fields with $E_y \ne B_z$.  For $x > 4000$ one has just the transmitted wave in dielectric $n_2 = 2$ with $n_2 E_y = B_z$.  When the fields are not equal, $E_y$ is in blue, $B_z$ is in red.}
}
\end{center}
\vspace{0cm} 
\end{figure}

\begin{figure}[!] \ 
\begin{center}
\vspace{0cm} 
\includegraphics[width=3.2in]{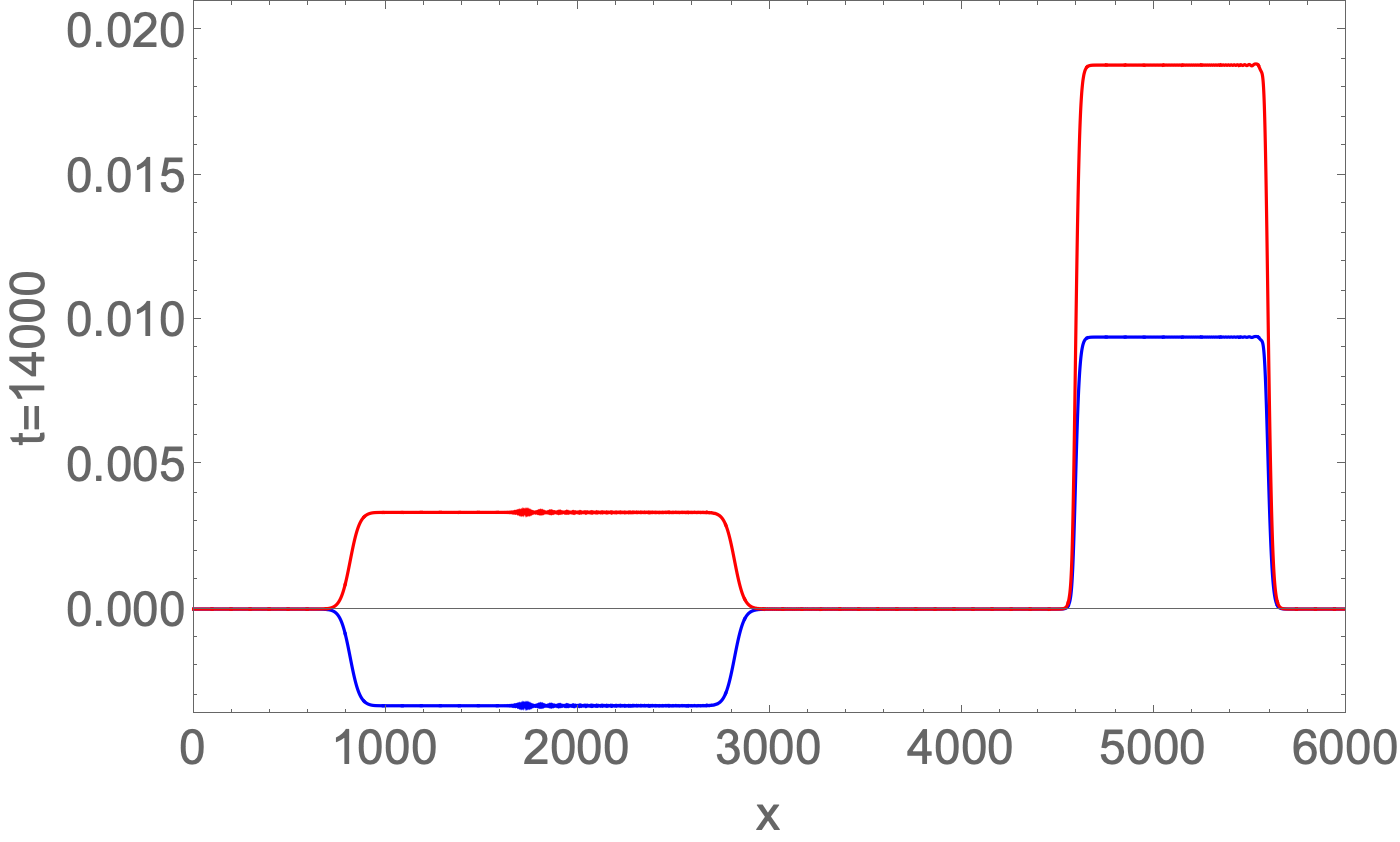}
\includegraphics[width=3.2in]{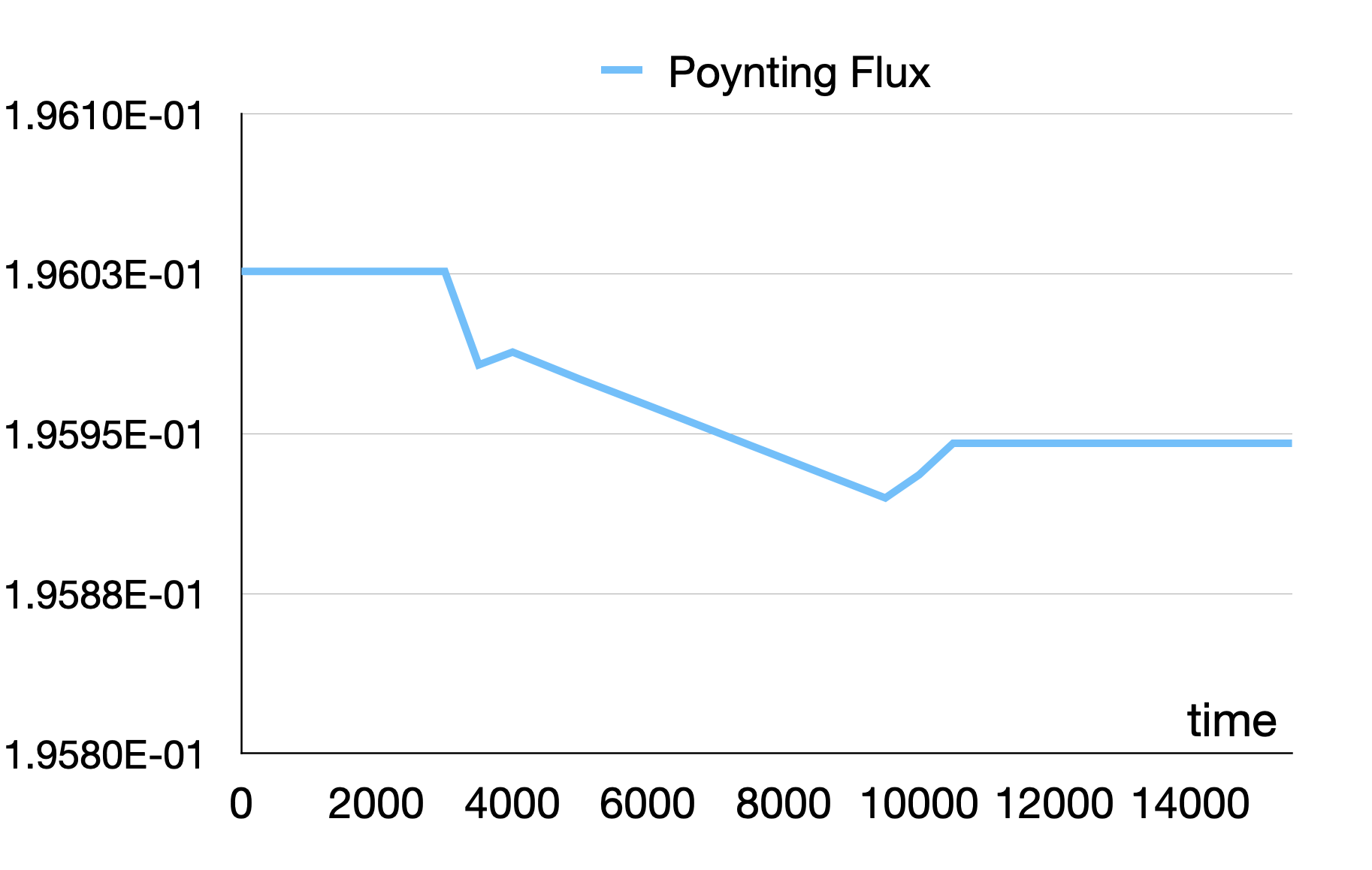}
\textcolor{red}{\textcolor{black}{(a)  quasi-asymptotic fields $E_y$, $B_z$  \quad   \quad \quad   \quad  \qquad  (b)  Poynting Flux.} }
\caption{\textcolor{black}{(a) The quasi-asymptotic state at $t=14000$ with the reflected pulse totally within medium $n_1$ and the transmitted pulse within region $n_2$.  $E_{refl}/E_{inc}$ = -0.334 = $(n_1-n_2)/(n_1+n_2)$, but $E_{trans}/E_{inc}$ = 0.941 = $2 \sqrt{n_1 n_2}/(n_1+n_2)$, since $n_1=1$ and $n_2=2$.  For the plane wave boundary value problem, the corresponding Fresnel ratio = $2 n_1/(n_1+n_2) = 0.667$
(b)  The Poynting flux as a function of time.  For $0< t < 3500$, the pulse is totally within the vacuum region $n_1$, while for $10500 < t < 17000$ the pulse is in its quasi-asymptotic state, with no overlap across the dielectric boundary layer.  For time $3500 < t < 10000$ the pulse is in a transient state, straddling both regions (c.f., Fig. 15b at $t = 6000$), and the straightforward quasi-asymptotic state evaluation of S(t) must be replaced by a more careful analysis, decomposing the pulse into incident and reflected parts.}
}
\end{center}
\vspace{0cm} 
\end{figure} 


\subsection{\textcolor{black}{QLA as a quantum algorithm}}
{\textcolor{black}{The initial conditions on the electromagnetic fields will be encoded onto the chosen qubit basis at each lattice node.
The unitary collision operators (e.g., Eq. (18) ) are composed of groupings of $2 \cross 2$ rotation matrices that entangle  pairs of qubits 
\begin{equation}
    \begin{bmatrix}
    \cos\theta  &  \sin\theta \\
    - \sin\theta  &  \cos\theta   
    \end{bmatrix}
\end{equation}
For example, 
for a 2-qubit state, we can represent its qubit basis as $ (| 00 \rangle \;, | 01 \rangle \;, | 10 \rangle \; , |  11 \rangle \; )$.  
If one applied the  $2 \cross 2$ rotation matrix, Eq. (44), to the subset $ (| 01 \rangle \;, | 10 \rangle  )$, with $\theta = \pi/4$ then one of
the resulting elements would be the Bell state $ (| 01 \rangle + |10 \rangle )/ \sqrt 2 $ - which is a maximally entangled qubit state.
The streaming operators move this entanglement throughout the lattice.  It is these unitary operators that permit us to consider the
QLA as a quantum computing algorithm.}

{\textcolor{black}{Our current QLA requires the introduction of some potential operators after the sequence of interleaved unitary collide-stream 
operators are performed (e.g., Eq. (38)).  Typically one of these is Hermitian, rather than unitary.  There are attempts by some authors [30] to develop algorithms
on approximating a sparse Hermitian operator by unitary operators.  Koukoutsis et. al. [31] have proven that there does exist 
 a fully unitary Maxwell algorithm even for inhomogeneous tensor dielectric media.  This is an area of current research.
 Because our QLA will require
error correcting qubits with long coherence times, it will be some time before we will be able to make sufficiently long runs to obtain interesting
physics on quantum computers.}

{\textcolor{black}{It is interesting to note that  some time ago Yepez [32] had devised a 1D QLA for the solution of Burgers equation.  Since the 1D Burgers equation has just one scalar field, only two qubits are required per spatial node.  These QLA unitary collide-stream operators were then encoded onto an NMR quantum information processor [33].  Using a weakly inhomogeneous magnetic field across the NMR sample, the experimenters [33] achieved (in 2006) a spatial grid of 16 lattice nodes, each with their distinct resonant frequencies that could be distinguished and modulated by shaped RF pulses.  Their coherence time permitted the evolution of the QLA for 8 time steps -- sufficient to see the start of the steepening of an initial oscillatory field towards a shock structure.}

\subsection{\textcolor{black}{QLA's parallelization on classical supercomputers}}
{\textcolor{black}{From the view point of parallelization of the QLA on a classical supercomputer, our algorithms [4, 12 - 16, 18, 19] have basically seen no saturation with the number of cores available.  Initially these 1D QLA's were benchmarked against exact soliton-soliton collision collisions for the integrable Korteweg- de Vries (KdV) and Nonlinear Schrodinger equation (NLS).  Using the modular structure of these QLA's we studied the non-integrable 2D and 3D NLS equations which, in condensed matter, are known as the Gross-Pitaevski (GP)-equation.  
There are indications that the QLA outperforms conventional standard computational fluid dynamic (CFD) codes.  For example, in the study of 3D quantum turbulence, it was conventional to introduce a wave-number-dependent dissipative term to damp out troublesome short wavelength perturbations [34].  This, of course, destroyed the Hamiltonian structure of the GP equation so that not only was the total energy not conserved, but also the total number.  The decay in the total number was obviated by the addition of a time-dependent chemical potential into the GP equation, but the total energy still decayed in time.   In our QLA [4] representation of quantum turbulence, the Hamiltonian structure was fully preserved and, on just under $12,000$ cores, we performed 3D quantum turbulence production runs on $5760^3$ grids.}

{\textcolor{black}{Finally, we present some 2015 timings on the IBM BlueGene/Q $MIRA$ at the Argonne National Laboratory Supercomputer Center.  As the collision operators are purely local and the streaming operators are simple shifts, QLA is ideally parallelized on classical supercomputers.  We [35] achieved a performance of 1.17 PetaFlops using $2/3^{rds}$ of $MIRA$ at a parallel efficiency of just over 94$\%$.  The code was optimized for OpenMP and strong scaling was performed on a fixed grid of $5120^3$ as we ramped up the number of cores from $65,536$ to $524,288$.  The code achieved over 36 GFlops/node.}


\section{References}
\quad [1]  YEPEZ, J. 2002 An efficient and accurate quantum algorithm for the Dirac equation. arXiv: 0210093. 

[2]  YEPEZ, J.  2005 Relativistic Path Integral as a Lattice-Based Quantum Algorithm. Quant. Info. Proc. 4, 471-509. 

[3]  YEPEZ, J, VAHALA, G $\&$ VAHALA, L.  2009a  Vortex-antivortex pair in a Bose-Einstein condensate, Quantum lattice gas model of   theory in the mean-field approximation.  Euro. Phys. J. Special Topics 171, 9-14

[4]  YEPEZ, J, VAHALA, G, VAHALA, L $\&$ SOE, M.   2009b Superfluid turbulence from quantum Kelvin wave to classical Kolmogorov cascades.  Phys. Rev. Lett. 103, 084501. 

[5]  YEPEZ, J. 2016 Quantum lattice gas algorithmic representation of gauge field theory. SPIE 9996, paper 9996-22

[6]  OGANESOV, A, VAHALA, G, VAHALA, L, YEPEZ, J $\&$ SOE, M.  2016a. Benchmarking the Dirac-generated unitary lattice qubit collision-stream algorithm for 1D vector Manakov soliton collisions.  Computers Math. with Applic.  72, 386

[7]  OGANESOV, A, FLINT, C, VAHALA, G, VAHALA, L, YEPEZ, J $\&$ SOE, M  2016b  Imaginary time integration method using a quantum lattice gas approach.  Rad Effects Defects Solids 171, 96-102

[8]  OGANESOV, A, VAHALA, G, VAHALA, L $\&$ SOE, M.  2018. Effects of Fourier Transform on the streaming in quantum lattice gas algorithms.  Rad. Eff. Def. Solids, 173, 169-174

[9]  VAHALA, G, VAHALA, L $\&$ YEPEZ, J.  2003 Quantum lattice gas representation of some classical solitons. Phys. Lett A310, 187-196

[10]  VAHALA, G, VAHALA, L $\&$ YEPEZ, J.  2004.  Inelastic vector soliton collisions: a lattice-based quantum representation. Phil. Trans: Mathematical, Physical and Engineering Sciences, The Royal Society, 362, 1677-1690

[11]  VAHALA, G, VAHALA, L $\&$ YEPEZ, J.  2005  Quantum lattice representations for vector solitons in external potentials. Physica A362, 215-221.

[12]  VAHALA, G, YEPEZ, J, VAHALA, L, SOE, M, ZHANG, B, $\&$ ZIEGELER, S. 2011 Poincaré recurrence and spectral cascades in three-dimensional quantum turbulence.  Phys. Rev. E84, 046713

[13]  VAHALA, G, YEPEZ, J, VAHALA, L $\&$SOE, M, 2012  Unitary qubit lattice simulations of complex vortex structures.  Comput. Sci. Discovery 5, 014013

[14]  VAHALA, G, ZHANG, B, YEPEZ, J, VAHALA. L $\&$ SOE, M.  2012 Unitary Qubit Lattice Gas Representation of 2D and 3D Quantum Turbulence.  Chpt. 11 (pp. 239 - 272), in Advanced Fluid Dynamics, ed. H. W. Oh, (InTech Publishers, Croatia)

[15]  VAHALA, G, VAHALA, L $\&$ SOE, M.  2020. Qubit Unitary Lattice Algorithm for Spin-2 Bose Einstein Condensates: I – Theory and Pade Initial Conditions.  Rad. Eff. Def. Solids 175, 102-112

[16]  VAHALA, G, SOE, M $\&$ VAHALA, L.  2020  Qubit Unitary Lattice Algorithm for Spin-2 Bose Einstein Condensates: II – Vortex Reconnection Simulations and non-Abelian Vortices.  Rad. Eff. Def. Solids 175, 113-119

[17]  VAHALA, G, VAHALA, L,  SOE, M $\&$ RAM, A, K.  2020.  Unitary Quantum Lattice Simulations for Maxwell Equations in Vacuum and in Dielectric Media, J. Plasma Phys $\bf{86}$, 905860518 

[18]  VAHALA, L, VAHALA, G $\&$ YEPEZ, J. 2003  Lattice Boltzmann and quantum lattice gas representations of one-dimensional magnetohydrodynamic turbulence. Phys. Lett  A306, 227-234.

[19]  VAHALA, L, SOE, M, VAHALA, G $\&$ YEPEZ, J.  2019a. Unitary qubit lattice algorithms for spin-1 Bose-Einstein condensates.  Rad Eff. Def. Solids 174, 46-55

[20]  VAHALA, L, VAHALA, G, SOE, M, RAM, A $\&$ YEPEZ, J.  2019b. Unitary qubit lattice algorithm for three-dimensional vortex solitons in hyperbolic self-defocusing media.  Commun Nonlinear Sci  Numer Simulat 75, 152-159 

[21]  DODIN, I. Y. $\&$ STARTEV, E. A. 2021 On appliations of quantum computing to plasma simulations.  Phys. Plasmas $\bf{28}$, 092101

[22]  ENGEL, A., SMITH, G $\&$ PARKER, S. E. 2019  Quantum Algorithm for the Vlasov Equation, Phys. Rev. $\bf{100}$, 062315

[23] ENGEL, A., SMITH, G $\&$ PARKER, S. E.  2021.  Linear Embedding of nonlinear dynamial systems and prospects for efficient quantum algorithms,  Phys. Plasmas $\bf{28}$, 062305

[24] LIU, J-P, KOLDEN, H.O, KROVI, H. K, LOUREIRO, N. F, TRIVISA< K, $\&$ CHILDS, A. M. 2021.  Proc. Natl. Acad. Sciences $\bf{118}$, e2026805118.

[25]  LAPORTE, O. $\&$ UHLENBECK, G. E. 1931 Application of spinor analysis to the Maxwell and Dirac equations. Phys. Rev. 37, 1380-1397.

[26]  OPPENHEIMER, J. R. 1931 Note on light quanta and the electromagnetic field. Phys. Rev. 38, 725-746.

[27]  MOSES, E.  1959  Solutions of Maxwell’s equations in terms of a spinor notation:  the direct and inverse problems,  Phys. Rev. 113, 1670-1679

[28]  KHAN, S. A. 2005  Maxwell Optics:  I.  An exact matrix representation of the Maxwell equations in a medium.  Physica Scripta 71, 440-442;  also arXiv: 0205083v1 (2002)

[29]  RAM, A. K., VAHALA, G., VAHALA, L. $\&$ SOE, M 2021  Reflection and transmission of electromagnetic pulses at a planar dielectric interface - theory and quantum lattice simulations AIP Advances 11, 105116 (1-12).

[30]  CHILDS, A, M $\&$ WIEBE, N.  2012.  Hamiltonian simulation using linear combinations of unitary operations.  Quantum Info. Comput.12, 901–924.

\textcolor{black}{[31] KOUKOUTSIS, E., HIZANIDIS, K., RAM, A. K., $\&$ VAHALA, G.  2022.  Dyson MAps and Unitary Evolution for Maxwell Equations in Tensor
Dielectric Media.  arXiv:2209.08523}

\textcolor{black}{[32]  YEPEZ, J., 2002.  An efficient quantum algorithm for the one-dimensional Burgers equation,  arXiv:0210092}

\textcolor{black}{[33]  CHEN, Z., YEPEZ, J., $\$$ CORY, D. G., 2006.  Simulation of the Burgers equation by NMR quantum information processing.  Phys. Rev. A74, 042321 (arXiv:0410198v1).}

\textcolor{black}{[34] KOBAYASHI, M. $\&$ TSUBOTA, M., 2005 Kolmogorov Spectrum of Superfluid Turbuelnce:  Numerical Analysis of the Gross-Pitaevskii Equation with a Small-Scale Dissipation.  Phys. Rev. Lett. 94, 065302.}

\textcolor{black}{[34] WILLIAMS, T. J.,  private communication and help in optimizing our QLA spin-1 Bose-Einstein Condensate code.}
\end{document}